\documentclass[12pt,preprint]{aastex}
\slugcomment{To appear in the ApJ Supplement Series}

\def\Vo{\ifmmode V_0\else $V_0$\fi}
\def\VCO{\ifmmode {\rm V_{CO}}\else V$_{\rm CO}$\fi}
\def\Ho{\ifmmode H_0\else $H_0$\fi}

\def\H2{\ifmmode {\rm H_2}\else H$_2$\fi}
\def\Ha{\ifmmode H\alpha\else H$\alpha$\fi}
\def\Hb{\ifmmode H\beta\else H$\beta$\fi}
\def\Brg{\ifmmode Br\gamma\else Br$\gamma$\fi}
\def\LB{\ifmmode L_B\else $L_B$\fi}
\def\Lsixty{\ifmmode L_{60}\else $L_{60}$\fi}
\def\LIR{\ifmmode L_{\rm IR}\else $L_{\rm IR}$\fi}
\def\LFIR{\ifmmode L_{\rm FIR}\else $L_{\rm FIR}$\fi}
\def\LHa{\ifmmode L_{\rm H\alpha}\else $L_{\rm H\alpha}$\fi}
\def\LCO{\ifmmode L_{\rm CO}\else $L_{\rm CO}$\fi}
\def\MHI{\ifmmode M_{\rm HI}\else $M_{\rm HI}$\fi}
\def\MH2{\ifmmode M_{\rm H_2}\else $M_{\rm H_2}$\fi}
\def\ICO{\ifmmode I_{\rm CO}\else $I_{\rm CO}$\fi}
\def\SCO{\ifmmode S_{\rm CO}\else $S_{\rm CO}$\fi}
\def\SHI{\ifmmode S_{\rm HI}\else $S_{\rm HI}$\fi}
\def\FHa{\ifmmode F_{\rm H\alpha}\else $F_{\rm H\alpha}$\fi}
\def\FIR{\ifmmode F_{\rm IR}\else $F_{\rm IR}$\fi}
\def\ftwelve{\ifmmode f_{12}\else $f_{12}$\fi}
\def\ftwenty{\ifmmode f_{25}\else $f_{25}$\fi}
\def\fsixty{\ifmmode f_{60}\else $f_{60}$\fi}
\def\fhundred{\ifmmode f_{100}\else $f_{100}$\fi}
\def\Tdust{\ifmmode T_{\rm dust}\else $T_{\rm dust}$\fi}
\def\WHa{\ifmmode W_{\lambda}(\Ha)\else $W_{\lambda}$(\Ha)\fi}
\def\Lsun{\ifmmode L_{\sun}\else $L_{\sun}$\fi}
\def\Msun{\ifmmode M_{\sun}\else $M_{\sun}$\fi}
\def\IRAS{{\it IRAS\/}}
\def\Bt{\ifmmode B_{\rm T}\else $B_{\rm T}$\fi}
\def\Ab{\ifmmode A_B\else $A_B$\fi}
\def\Av{\ifmmode A_V\else $A_V$\fi}
\def\kms{\ifmmode {\rm km~s^{-1}}\else km~s$^{-1}$\fi}
\def\um{\ifmmode \mu m\else $\mu$m\fi}
\def\Vband{\ifmmode V{\rm -band}\else $V$-band\fi}
\def\Rband{\ifmmode R{\rm -band}\else $R$-band\fi}
\def\Iband{\ifmmode I{\rm -band}\else $I$-band\fi}
\def\Hband{\ifmmode H{\rm -band}\else $H$-band\fi}

\def\nod{\nodata}
\def\pl{\phm{$<$}}

\shortauthors{Bushouse et al.}
\shorttitle{Near-IR Atlas of ULIRGs}
 
\begin{document}

\title{Ultraluminous Infrared Galaxies: Atlas of Near-Infrared Images
\footnote{
Based on observations with the NASA/ESA {\it Hubble Space Telescope},
obtained at the Space Telescope Science Institute, which is operated by the
Association of Universities for Research in Astronomy, Inc., under NASA
contract NAS5-26555.} }

\author{H.~A.~Bushouse\altaffilmark{2},
	K.~D.~Borne\altaffilmark{3},
	L.~Colina\altaffilmark{4},
	R.~A.~Lucas\altaffilmark{2},
	M.~Rowan-Robinson\altaffilmark{5},
	A.~C.~Baker\altaffilmark{6},
	D.~L.~Clements\altaffilmark{6},
	A.~Lawrence\altaffilmark{7}, and
	S.~Oliver\altaffilmark{8}
 }

\altaffiltext{2}{Space Telescope Science Institute, 3700 San Martin Drive,
 Baltimore, MD 21218; bushouse@stsci.edu, lucas@stsci.edu}

\altaffiltext{3}{Raytheon Information Technology and Scientific Services,
 NASA Goddard Space Flight Center, Greenbelt, MD 20771;
 borne@rings.gsfc.nasa.gov}

\altaffiltext{4}{Instituto de Estructura de la Materia,
 Consejo Superior de Investigaciones Cientificas, Serrano 121,
 28006 Madrid, Spain; colina@isis.iem.csic.es}

\altaffiltext{5}{Astrophysics Group, Blackett Laboratory, Imperial College,
 Prince Consort Road, London SW7 2BW, UK;
 m.rrobinson@ic.ac.uk}

\altaffiltext{6}{University of Wales, College of Cardiff, Physics and
 Astronomy Department, P.O. Box 913, Cardiff CF2 3YB, UK;
 a.baker@astro.cf.ac.uk, david.clements@astro.cf.ac.uk}

\altaffiltext{7}{Institute for Astronomy, University of Edinburgh, Royal
 Observatory, Blackford Hill, Edinburgh EH9 3HJ, UK; a.lawrence@roe.ac.uk}

\altaffiltext{8}{Astronomy Centre, University of Sussex, Falmer, Brighton
 BN1 9QJ, UK; s.oliver@sussex.ac.uk}

\begin{abstract}
A sample of 27 ultraluminous infrared galaxy (ULIRG) systems has been
imaged at 1.6~\um\ using the HST Near Infrared Camera and Multi-Object
Spectrometer (NICMOS).
These ULIRGs are from a larger sample also imaged with HST in the \Iband.
Images and catalog information for the NICMOS subsample, as well as brief
morphological descriptions of each system are presented.
Inspection of the infrared images and a comparison with optical images of
these systems shows that at least 85\%\ are obviously composed of two or more
galaxies involved in a close interaction or merger event, with as many as
93\%\ showing some signs of interaction history.
Approximately 37\%\ of the systems show either spectroscopic or morphological
characteristics of an active galactic nucleus (AGN).
The infrared morphologies of these systems are generally less complicated
or disturbed than their optical morphologies, indicating that some of the
small-scale features seen in optical images are likely due to 
complicated patterns of dust obscuration, as well as widely distributed star
formation activity.
In some systems the high-resolution HST infrared images have revealed nuclear
remnants that are obscured or unidentified in ground-based imaging,
which has led to changes in previously determined interaction
stage classifications or system content.
In general, however, the NICMOS images support previous conclusions from
previous HST optical imaging.
\end{abstract}


\section{Introduction}

One of the major discoveries of the {\it Infrared Astronomical Satellite}
\footnote{The {\it Infrared Astronomical Satellite} was developed and
operated by the US National Aeronautics and Space Administration (NASA), the
Netherlands Agency for Aerospace Programs (NIVR), and the UK Science and
Engineering Research Council (SERC).}
(\IRAS) mission was the existence of a population of galaxies that emit
the majority of their luminosity in the far-infrared \citep[e.g.][]{soi84}.
At luminosities greater than $\LIR(8-1000~\um)>10^{11}\Lsun$,
infrared-selected galaxies become more numerous than optically-selected
starburst and
Seyfert galaxies of comparable bolometric luminosity, and at the highest
luminosities, $\LIR>10^{12}\Lsun$, they exceed the space densities of
quasi-stellar objects (QSOs) by a factor of 1.5--2 \citep{soi87,san89}.

Subsequent studies have shown that a large fraction of ultraluminous IR
galaxies (ULIRGs), with $\LIR>10^{12}\Lsun$, are interacting or merging
systems, and many contain active galactic nuclei (AGN)
\citep[for a review see][]{san96}.
The relative fractions of systems that are interacting or contain AGN,
however, varies depending on the particular subsample of ULIRGs studied.
Some samples appear to contain only 10--20\%\ interacting, merging, or
peculiar systems, while in others the fraction is 80\%\ or higher 
\citep[e.g.][]{san88a,mel90,zhe91,lee94,kim95,cle96b,mur96}.
Similarly, the fraction of systems containing AGN varies from 
$\sim$25\%\ \citep[e.g.][]{lee89,law99,vei95,vei99b} to nearly
100\%\ \citep[e.g.][]{san88a,san00}.
In general, the fraction of interacting systems and systems containing AGN
appears to increase with IR luminosity \citep{kla89,kla93,row91,kim98,vei99a,
vei99b}.

Given the propensity for ULIRGs to be found in interacting systems, it
seems clear that collision and merger driven processes are the major
mechanisms responsible for the bulk of their infrared emission.
Starbursts, induced by the gravitational interaction of colliding and merging
galaxies, could account for the excess IR emission
\citep[e.g.][]{lee89,ash95,lut98,gen98,lut99}.
On the other hand, the frequent occurence of AGN in these systems suggests
that non-thermal processes might be the dominant energy source in the most
luminous systems \citep[e.g.][]{san88a,vei95,gen98}.

Recent deep surveys at 850~\um\ using the SCUBA camera
\citep[Submillimeter Common User Bolometer Array;][]{hol99}
on the James Clerk Maxwell Telescope have discovered a substantial population
of ULIRGs at high redshift ($z\sim$1--4), whose cumulative space density
accounts for nearly all of the extragalactic background light at
submillimeter wavelengths \citep{sma97,hug98,bar98,bar99,eal99,bla99}.
Evidence suggests that the SCUBA sources, like local ULIRGs, are powered by
intense starbursts and AGN, both fueled by mergers of gas-rich disk galaxies.
As such, they represent a key stage in the early evolutionary
history of all galaxies, and are likely to be the precursors of QSOs.
Studies of low-$z$ ULIRGs, which appear to be local templates of the SCUBA
sources, will therefore lead to a greater understanding of massive star
formation, the formation and evolution of galaxies and QSOs, as well as the
metal enrichment of the intergalactic medium.

Recent Hubble Space Telescope (HST) and other high-resolution imaging
observations of small and
medium-sized samples of ULIRGs have demonstrated the ability to resolve
close nuclei, near-nuclear star clusters, and the complicated
morphologies that are present at all scales in these types of systems.
\citet{boy96} and \citet{sur00}, for example, have found systems with close
double nuclei that were previously unresolved.
\citet{sur98}, \citet{zhe99}, and \citet{sco00} have utilized HST images
to measure the surface brightness profiles in the inner regions of ULIRGs,
and found that many are well-fit by an $R^{1/4}$ law, indicating that the
merging events may be in the process of turning these systems in giant
elliptical galaxies.

The study of a larger, more representative collection of ULIRGs is
needed to address with strong statistical significance both the
``interaction frequency'' and ``starburst vs.~AGN'' issues mentioned above.
We are carrying out a multi-wavelength HST imaging survey of a large,
well-defined sample of ULIRGs, in order to survey the fine-scale features that
are associated with the interaction and activity-related processes that are at
work within these systems. 
\citet{bor01} present an atlas of Wide Field Planetary Camera-2 (WFPC2)
\Iband\ images of the parent sample of 123 ULIRG systems.
\citet{far01} present the results of an additional WFPC2 \Vband\ survey of a
subsample of 23 of these systems.

Here we present near-IR survey results, consisting of Near Infrared Camera
and Multi-Object Spectrometer (NICMOS) \Hband\ images of a representative
subsample of 27 ULIRGs.
Additional wavelength coverage at 4--200~\um\ from Infrared Space Observatory
(ISO) observations has also been obtained for this subsample.
Images and catalog information for the NICMOS sample, as well as brief
morphological descriptions of each system, are contained in this paper.
\citet{col01} present more detailed analyses of these data.
The near-IR imaging data complement the optical data in several ways.
The near-IR band is less sensitive to emission from young, blue star
clusters, as well as the absorption effects of interstellar dust. A comparison
of the near-IR and optical images allows us to ascertain to what level the
complex small-scale morphological structures seen in the optical images are
due to either dust or young star clusters.
This will provide confirmation of conclusions previously based on the optical
morphology alone regarding the makeup and status of the ULIRGs.
Futhermore, the addition of the NICMOS data will yield $I-H$ colors, which
can be used, for example, to classify various morphological features as
remnant nuclei or young star clusters.

\section{The Near-IR ULIRG Sample}

The sample of ULIRGs imaged in the near-IR was randomly selected from the
larger sample of 123 systems previously imaged in the \Iband\ with WFPC2.
Full details of the selection criteria for the WFPC2 sample are given by
\citet{bor01}.
It consists of objects compiled by \citet{san88a} and \citet{san88b}, southern
hemisphere targets from \citet{mel90},
many systems from the QDOT all-sky \IRAS\ galaxy redshift survey by
\citet{law99}, and complemented with systems from \citet{lee94},
\citet{kimetal95}, and \citet{cle96a}.
The near-IR subsample presented in this paper was compiled by
selecting a representative subsample of 50 targets from the WFPC2 sample,
consisting mainly of galaxies from the QDOT survey, as well as the samples of
\citet{mel90} and \citet{cle96a}.
The objects were selected without regard for morphological or spectroscopic
types.
HST NICMOS snapshot images were obtained for a total of 27 systems in this
subsample, of which 17 are from the QDOT survey, 9 from the \citet{mel90}
sample, and 1 from the \citet{cle96a} sample.

Table~\ref{tbl1} lists the objects that have been observed with NICMOS and
includes basic catalog information for each object.
Redshifts and \IRAS\ fluxes were obtained for each object from the NASA/IPAC
Extragalactic Database (NED), where the \IRAS\ fluxes are from the Faint
Source Catalog \citep{mos90}.
This sample covers a redshift range of $0.05<z<0.30$, with a median
$z=0.17$.
The vast majority (85\%) of the systems lie within a redshift range of
$0.08<z<0.23$.
The 60\um\ luminosities range from $10^{11.69}$ to $10^{12.46}\Lsun$,
with a median $\Lsixty=10^{11.99}\Lsun$.
Most of the galaxies (81\%) have infrared colors indicative of ``cool''
ULIRGs (i.e. $f_{25\um}/f_{60\um}<0.2$).
The distribution of 60\um\ luminosities is essentially random within the
redshift range $0.08<z<0.21$.
These systems therefore form a representative sample of low-redshift
cool ULIRG's.

\citet{san88a} originally defined ULIRGs as having
$\LIR>10^{12}\Lsun$, where the IR luminosity is integrated over the
8--1000~\um\ range and makes use of flux measurements from all four
\IRAS\ bands.
Very few of the objects in the current sample, however, have detections in the
\IRAS\ 12 and 25~\um\ bands, therefore it is not possible to derive
comparable IR luminosities.
Furthermore, some objects in this sample also do not have detections in the
\IRAS\ 100~\um\ band, making it similarly impossible to derive the typical
far-infrared luminosity, \LFIR, from the 60 and 100~\um\ flux densitites
\citep[e.g.][]{ful89} in a uniform fashion.
Table~\ref{tbl1} therefore lists only the 60\um\ luminosity for each object.
For the types of galaxies in this sample, \Lsixty\ is roughly equivalent to
\LFIR, and $L_{[8-1000~\um]}$ values are roughly 2\LFIR\ for a wide range of
dust temperatures and emissivities \citep{law89}. 
Therefore a definition of ultraluminous which is equivalent to that of
\citet{san88a} is $\Lsixty>10^{11.7}\Lsun$.
All of the objects in the current sample meet this criterion.

\section{Observations and Data Reduction}

Direct imaging observations of 27 ULIRGs were obtained with the NICMOS
instrument on HST in a snapshot program that executed from January to
August 1998.
The NICMOS camera 2 was used, which has a pixel size of
$\sim$0.075\arcsec\ and a field-of-view of $\sim$19.2\arcsec\ square.
The images were taken with the NICMOS F160W filter, which approximates the
groundbased \Hband.
Observations of each target were obtained using a four-point square dither
pattern, with dither steps of 3.75\arcsec\ ($\sim$50 pixels) between each
exposure.
Each of the four images for a given target was taken in the NICMOS 
multiaccumulate (or ``multiaccum'') mode of operation, using the ``STEP32''
readout sequence and 12 readout samples, for a total integration time per
exposure of 160~s.
Combining the four exposures for each target yields an image with a
total field-of-view of $\sim$23\arcsec\ square and a total on-source
integration time of 640~s.

The raw data files were processed with the IRAF/STSDAS NICMOS
calibration program {\tt calnica} \citep{bus97}.
This program performs the operations necessary to subtract the initial 
(or ``zeroth'') readout of the multiaccum exposure sequence from all
subsequent readouts, flag hot and cold pixels, subtract the dark current,
linearize the detector photometric response, and flatfield the data.
This program also combines the multiple readouts of each multiaccum exposure
into a single image, and in the process rejects samples suspected of
containing cosmic-ray hits.

Two characteristics of the NICMOS detectors resulted in effects that were
severe enough in these data to require special processing.
First, there is a bias-related signal, known as ``shading'', that varies
across each image quadrant as the pixels are read out.
The exact shape and amplitude of the shading are dependent on
both detector temperature and the elapsed time since the previous readout.
Because the temperature characteristics of the standard dark current images used
in {\tt calnica} were a bit different than that of the science
data, the shading was not completely removed by subtracting the dark images.
The residual shading appears as a signal gradient across each quadrant in
a calibrated image.
To reduce these effects, the raw data were reprocessed with version 3.3 of
{\tt calnica}, which allows for the rejection of data from several of the
initial readouts of each exposure sequence in the processing step where the
data are combined to form a single image.
Rejection of the early readouts, which have the most non-linear increases
in exposure time from readout to readout, dramatically reduces the level of
the residual shading in the final combined image, while sacrificing very
little of the total integration time.

Second, the overall bias level in the NICMOS detectors is also
temperature-dependent, such that a residual bias signal
(often referred to as ``pedestal'') is frequently present in processed data.
The amplitude of the residual bias varies from quadrant to quadrant in each
image, and the process of flatfielding the data leaves a negative imprint of
the flatfield image, resulting in a non-uniform
background in the calibrated images.
To remove this effect, the calibrated data were processed with the
IRAF/STSDAS NICMOS program {\tt pedsky} \citep{bus00},
which iteratively solves for the
best combination of sky background and residual bias signal levels that
minimizes the flatfield residuals in each image.
The computed sky and bias (or ``pedestal'') signals are subtracted from the
images by this process.

After applying these processing steps to the four images acquired
for each target, the associated sets of calibrated images were then processed
with the IRAF/STSDAS NICMOS program {\tt calnicb}
\citep{bus97}.
This program combines the four dithered images into a final mosaiced image,
using a cross-correlation technique to determine the relative registrations
and a simple bilinear interpolation for accounting for subpixel shifts.
Figure~\ref{fig-images} shows the final combined images for the 27 ULIRGs
that have been observed in this program.
This figure includes contour plots and greyscale images of the NICMOS data
for each system, and a WFPC2 \Iband\ image for comparison.

Global \Hband\ absolute magnitudes for each system are listed in
Table~\ref{tbl1}.
The photometric measurements were performed on the final combined and
calibrated NICMOS images.
The measured count rates have been converted to apparent \Hband\ magnitudes
in the Vega system using the expression
\begin{equation}
m_H = -2.5 \times log(PHOTFNU \times countrate / ZP(Vega))
\end{equation}
where $PHOTFNU=2.07006\times 10^{-6}$ Jy$\cdot$sec/DN,
countrate is the integrated count rate, in units of DN/sec, and
$ZP(Vega)=1113$ Jy.
The resulting magnitudes have an estimated systematic uncertainty of
less than 5\%.
The majority of these measurements were performed using an aperture of
7.5\arcsec\ in diameter.
Larger apertures were used for a few particularly extended systems.
See \citet{col01} for more details.

\section{General Results}

The WFPC2 \Iband\ imaging survey of the parent sample of ULIRGs 
\citep{bor01} has revealed fine structure on all spatial scales in these
systems.
Only about 10\%\ of the systems have structure that is smooth and
centrally concentrated.
The rich variety of morphological features seen at optical wavelengths
is almost certainly due to a combination of tidal disturbances, young
star-forming regions, and dust obscuration.
Furthermore, the high spatial resolution of the WFPC2 images has revealed
the presence of multiple nuclei and other morphological distortions in systems
that were previously classified
as non-interacting on the basis of ground-based imaging.
There is also a strong correlation between the presence of
spatially-unresolved nuclear morphology (i.e. bright, point-like nuclei)
in the \Iband\ images of these systems and spectral classification as AGN.
This type of morphology therefore appears to be a strong
indicator for the presence of visible AGN (deeply obscured AGN could still be
present in systems that don't show this type of morphology).

The NICMOS \Hband\ survey images presented here show similar trends.
Table~\ref{tbl2} lists the global and nuclear morphologies,
indicating the number of discernable galaxies and nuclei, as well as the
presence of an unresolved nuclear source in each system.
Out of the 27 systems imaged, 23 are obviously composed of strongly
interacting multiple galaxies, or single-body systems with extensive shells,
loops, or tails that indicate an advanced stage of merging.
The remaining four systems---IRAS~04413+2608, IRAS~05233-2334, IRAS~20176-4756,
and IRAS~20414-1651---show a single, dominant galaxy with at least one or
more smaller, nearby companions.
In the IRAS~20176-4756 system, the companion is very distorted and therefore
may have suffered a collision with the parent galaxy.
The parent galaxy in IRAS~20414-1651 appears quite disturbed in the
\Iband\ WFPC2 image, suggesting a previous history of either collisions or
a merger event.
The morphologies of the parent and companion galaxies in IRAS~04413+2608 and
IRAS~05233-2334 are quite regular, with no obvious signs of previous or
current interaction activity.
The parent galaxy in IRAS~04413+2608, however, does harbor a
Seyfert~2 nucleus.
Therefore at least 85\%\ of these systems are interacting, with
as many as 93\%\ showing some signs of interaction history.

A little more than one-third of this sample contains obvious signs of an AGN.
There are 7 systems that have a bright, point-like nucleus visible in the
NICMOS \Hband\ images, suggestive of the presence of an AGN.
Previous spectral observations of galaxies in this sample yield a total of
8 objects with indications of either Seyfert 1 or 2 characteristics, and an
additional four objects with LINER spectra \citep{vei95,duc97,law99}.
Of these 12 objects, five show a bright, unresolved nuclear source in the
NICMOS \Hband\ images.
All five of these are classified spectroscopically as Seyfert 1 or 2 systems.
None of the systems classified as LINER shows any morphological evidence
for a strong nuclear source.
Thus the total number of systems that have either morphological or
spectroscopic indications of Seyfert activity is 10, yielding a fractional
content of AGN in this sample of $\sim$37\%.
This result is similar to the 34\%\ fraction of Seyferts found in
an \IRAS\ 1 Jy sample of ULIRGs by \citet{kim98}.
These data, however, do not rule out the existence of heavily
obscured AGN in these systems.
Furthermore, 12 of the systems in this sample do not have spectral
observations available at this time.
So the derived fraction of systems containing AGN should be viewed as a
lower-limit.

The combination of high spatial resolution, low background, and reduced
effects of dust obscuration offered by the NICMOS images gives us new
insights into many of the ULIRG systems when compared with previously obtained
ground-based optical and near-IR images.
For example, IRAS~11095-0238 was previously classified as a system containing
a single nuclei, and therefore at the end stage of a merger \citep{duc97}.
The NICMOS image of this system shows two unmerged cores, with a
separation of $\sim$0.5\arcsec.
Similarly, IRAS~16541+5301 appears as a single object in ground-based R-band
images \citep{lee94}, while the NICMOS image clearly reveals two remnant
galaxies within a common envelope.
Furthermore, IRAS~18580+6527, IRAS~19297-0406, and IRAS~20100-4156 are all
systems in which single components that appear in ground-based images have
been resolved by NICMOS into two or more sub-components.

Comparison of the NICMOS images with the somewhat
shorter-wavelength WFPC2 images has also helped to confirm the character of
many other systems in the sample, whose optical morphologies are sometimes
complicated by the effects of dust and the presence of young, blue star
forming regions.
In the IRAS~05233-2334, IRAS~16455+4553, IRAS~20037-1547, IRAS~20087-0308,
IRAS~20176-4756, and IRAS~20414-1651 systems, for example, the near-nuclear
regions have very complicated optical morphologies, showing multiple
bright regions that might be identified as remnant nuclei.
In the NICMOS \Hband\ images of these systems, however, we clearly see
a single, dominant nuclear core.
Similarly, the inner regions of IRAS~03538-6432, IRAS~10558+3845,
IRAS~11095-0238, and IRAS~2320-6926 also have very complex optical
morphologies, but the \Hband\ images very clearly
show two distinct, dominant nuclear cores.

Two systems in this sample, IRAS~18580+6527 and IRAS~19297-0406, were
previoulsy identified by \citet{bor00} as multiple progenitor systems,
based on the appearance of multiple nuclear-like cores in WFPC2
\Iband\ images.
The NICMOS images of these two systems support this conclusion, as they
also contain evidence for multiple remnant nuclei.
Quantitative analysis, however, will be required to
determine whether the putative galaxy cores have luminosities and colors
consistent with that expected for galaxy bulges.

\section{Object Descriptions}

Here we describe the morphology of each of the 27 ULIRGs observed with
NICMOS, and compare it to that seen in the \Iband\ images obtained with
the WFPC2 \citep{bor01}, as well as ground-based images obtained by
other investigators.

%%% 01 - Visit 06
\subsection{IRAS 03538$-$6432}

This object appears to be a relatively advanced merger, with a single
central body containing multiple nuclei, one prominent tidal tail, and
several additional short tails. In our WFPC2 \Iband\ image, the central
region of the main body contains several obvious dust lane features, and
what appear to be as many as three stellar concentrations. The NICMOS
\Hband\ image clearly reveals that there are in fact only two remant nuclei.
The third bright concentration seen in the \Iband\ image appears to be a hot
spot located on an arm of one of the two galaxies. The dominant
tidal tail is just barely visible in the \Hband\ image.

%%% 02 - Visit 08
\subsection{IRAS 04413+2608 (Sy 2)}

This is a pair of galaxies made up of a (possibly barred) spiral and a much
smaller spheroidal companion. In the \Hband\ image the spiral galaxy is
dominated by nuclear point source, presumably corresponding to the Seyfert 2
component. In the \Iband\ image, the bar component of the spiral galaxy appears
to be one-sided, i.e. it is much brighter on one side of the nucleus than
the other. This asymmetry is also visible in the \Hband, although to a lesser
degree. This suggests that at least some of the imbalance in the surface
brightness of the bar is due to dust obscuration. In the \Hband\ image there
is still a clear separation between the nuclear component and the hot spot
that appears at the base of a spiral arm on the faint end of the bar. If this
bifurcation is due to dust obscuration, the fact that it is still visible at
\Hband\ indicates a substantial amount of dust. Alternatively, the hot spot
may be an intrinsic bright feature, such as a site of recent starburst activity.

%%% 03 - Visit 10
\subsection{IRAS 05233$-$2334}

This is a single, disturbed looking spiral galaxy, with a somewhat fainter
spiral companion, and a much smaller spheroidal companion. The very complex
\Iband\ appearance of the central region of the main spiral is replaced in
the \Hband\ by a much more normal looking two-armed spiral, dominated by a
point-like nucleus. There is only a hint of structure in the near-nuclear
region, with a hot spot on one side of the nucleus that shows up in both
the I and \Hband\ images. The one-armed spiral companion appears at edge of
the \Hband\ frame and is dominated by an unresolved nucleus.
The only evidence for an interaction history in this system is the
presence of the two nearby companions.

%%% 04 - Visit 12
\subsection{IRAS 06206$-$6315 (Sy 2)}

This is a spectacular pair of interacting spiral galaxies, still clearly
separated, and yet involved in a very strong interaction that has produced
one very large tidal tail, and possibly other smaller ones. In the
\Iband\ image the central regions show very complex structure, apparently due to
mixtures of dust obscuration and active star formation regions. Even the
dominant tidal tail shows structure on small scales. The \Hband\ image clearly
reveals the nuclei of the two progenitor galaxies, with a fairly bright linear
feature connecting the two. While the \Hband\ image shows much less complicated
structure than the \Iband\ (indicating that much of that structure is due to
dust and possibly groups of hot stars), a few of the hot spots are still
visible in the near IR. This is true not only in the bodies of the galaxies,
but also in the large tidal tail. If these hot spots are in fact regions of
interaction-induced star formation, then the burst ages must be large
enough for a population of red supergiants to exist and be visible in the
near-IR.
Ground-based optical and near-IR imaging of this system has been able to
resolve the two galaxies, but does not reveal the much smaller scale knots
and dust signatures within the tidal tail and around the nuclei
\citep{duc97,rig99}.

%%% 05 - Visit 13
\subsection{IRAS 06268+3509}

This is another pair of strongly interacting spirals, still clearly
separated, yet connected by a tidal bridge, and each galaxy has a large
tidal tail. In the \Iband\ image the complex structure in the central
regions of the galaxies makes it difficult to identify exactly the location
of the nuclei. The \Hband\ image, on the other hand, clearly shows the nucleus
of each galaxy. The southern galaxy (upper right in Fig.~\ref{fig-images})
also shows spiral (or possibly tidal?) arm structures on either side of the
central core.
The two prominent tidal tails are also visible in the near-IR.
In particular, the middle portion of the tail from the northern (lower left)
galaxy is quite bright, indicating that it is composed of a
substantially old stellar population. Two or three hot spots near the base of
the tidal tail of the southern galaxy are also still visible in the
\Hband\ image,
but the rest of the small-scale structure attributable to dust obscuration is
largely absent.

%%% 06 - Visit 14
\subsection{IRAS 06361$-$6217}

\citet{zhe91} described this system as containing several well separated
objects, with the main central object appearing as a single, undisturbed
galaxy.
The HST images, however, reveal that the central object appears to be
made up of a small, edge-on galaxy that is in the process
of plunging through the center of a more spheroidal galaxy. In the
\Iband\ image the spheroidal galaxy nucleus is visible, with hot spots on either
side that appear to form portions of a ring around the nucleus. One of these
hot spots is still visible in the \Hband\ image, but is insignificant next to
the extremely bright, point-like nucleus (the first airy ring of the
diffraction pattern is clearly visible). In the \Iband\ the edge-on galaxy
shows several bright condensations along the major axis (pointing towards
the spheroidal galaxy). In the near-IR only one of these hot spots is
visible, and may represent the nucleus.

%%% 07 - Visit 15
\subsection{IRAS 06561+1902}

The \Hband\ image of this small group of galaxies covers only the central
binary pair and is rather unremarkable. The extensive common envelope of
tidal debris surrounding the pair is not detected in the \Hband\ image, nor
is the object that appears to be a very faint companion located within the
common envelope.
There is little to no evidence of fine-scale structure within the two
main galaxies in either the I or \Hband\ images. The nucleus of the southern
galaxy (upper left in Fig.~\ref{fig-images})
shows a bright, point-source nucleus in the \Hband\ image
(the NICMOS PSF is quite distinct).

%%% 08 - Visit 16
\subsection{IRAS 07381+3215}

This system shows a small, single galaxy, with some evidence of past
interaction. The extended, low surface brightness features, such as
tails and shells, that constitute this evidence in the optical are
not visible in the \Hband\ image. Small scale features, such as arms and
hot spots, that are visible in the central regions of the galaxy in the
\Iband\ image are barely visible in the near-IR. The \Hband\ image shows a
single nucleus with what appear to be a central bar and two spiral
arms wrapping out into the disk of the galaxy. A couple of localized
hot spots that appear in the optical image along the arms are not
visible in the near-IR.

%%% 09 - Visit 18
\subsection{IRAS 10558+3845}

Optical images show a single central body with a pair of symmetric, fan-like
tails, suggesting a fairly advanced merger. The central region is also quite
complex in the optical image,
with indications of either multiple nuclei
or hot spots, as well as patches of dust obscuration. The NICMOS
\Hband\ image shows the same single, central body, but now there are two
distinct nuclear concentrations, with the one towards the east (bottom)
being very bright and point-like. This nucleus is essentially invisible
in the \Iband\ image, suggesting that it is hidden by large amounts of dust.
The western (upper) nuclear-like region is fainter and more extended, and the
ellipticity
suggests an edge-on orientation. Only the innermost (highest surface
brightness) portions of the fan-like tails are detected in the \Hband\ image.

%%% 10 - Visit 19
\subsection{IRAS 11095$-$0238 (LINER)}

Ground-based optical and near-IR images by \citet{duc97} shows two tidal
tails in this system, but only a single nucleus.
The HST \Iband\ image shows a single main body with two clearly resolved
cores, and a long, complex tidal tail. It is not obvious whether the two
brightest (and also closest) nuclear condensations are in fact multiple
nuclei or simply a single core bifurcated by a dense dust lane. The
NICMOS image of the core is significantly less complex, showing the same two
dominant nuclear concentrations and a few of the hot spots that are visible
further out in the optical image. But even in the \Hband\ image the edges of
the low intensity band that separates the two nuclei are sharp enough to
suggest that we still may be seeing a dust lane.

%%% 11 - Visit 21
\subsection{IRAS 13352+6402}

This is a small group of galaxies, dominated by a central, strongly interacting
pair.
A ground-based \Rband\ image of this system shows the pair of galaxies in close
contact, with some evidence for a tidal extension from one of them
\citep{lee94}.
In the HST images the northeastern (lower) galaxy of the pair appears to be a
nearly face-on spiral.
The arm pattern is very asymmetric, with the western arm being much brighter.
The southwestern (upper) galaxy is smaller, with a large tail containing
several condenstations (star forming complexes?).
There is a bright knot at the very
end of the tail. The \Hband\ image shows almost no small-scale features 
within these two galaxies, although the asymmetry of the arms in the spiral
is still noticeable. A few of the bright knots in the tail of the western
galaxy, most notably the one at the very end of the tail, are also still
visible in the near-IR.

%%% 12 - Visit 22
\subsection{IRAS 13469+5833}

The \citet{lee94} ground-based \Rband\ image of this system shows what is
probably two galaxies in contact, with two tidal tails that wrap around
the pair.
The HST images confirm that this is indeed a close pair of strongly interacting
spiral galaxies, in the late stages of collision, but not yet merging.
In the WFPC2 \Iband\ image the central region
of the eastern (lower right) galaxy is complex, with several hot spots and
indications of
patchy dust obscuration. The extended regions of the tail that lies to the
northeast of this galaxy
also have complex fine-scale structure, with numerous star complexes and
dust lanes. The central region of the western (upper) galaxy appears
smooth, with
a single nucleus, in both the I and \Hband\ images. The eastern galaxy
appears
less complex in the \Hband, but there is still a slight indication of a
double peak at the very center, with a separation of only $\sim0.2$\arcsec.
The bright ridge of emission seen along the edge of the tail from this
galaxy in the \Iband\ image is still visible in the near-IR, as is
the object (dwarf galaxy?) located within the inner arc of the tail from
the western galaxy.

%%% 13 - Visit 24
\subsection{IRAS 14378$-$3651 (Sy 2)}

Ground-based optical and near-IR images of this system show a group of
isolated objects, with a central elliptical-like galaxy that has no
remarkable features \citep{mel90,duc97}.
The HST images reveal that the main galaxy has multiple shells and an
inner spiral pattern, along with a faint companion to the south
(upper right).
The center of the main
galaxy appears to have a single nucleus in both the optical and near-IR 
HST images.
The nucleus in the \Hband\ image is point-like, showing the instrumental
diffraction spikes.
The shells and inner arms are faintly visible in the \Hband\ image, as are
two bright spots (star clusters?) that appear in the outer envelope.

%%% 14 - Visit 26
\subsection{IRAS 16159$-$0402}

This is a compact group of at least two and possibly three strongly interacting
galaxies.
The \citet{lee94} ground-based \Rband\ image shows two galaxies
in contact, but with no obvious tidal extensions or internal disturbances.
In the WFPC2 \Iband\ image the brightest galaxy of this pair has a complex
inner disk structure, which has the appearance of being due to dust lanes.
In the \Hband\ image some of this structure is still visible, indicating large
amounts of dust.
The single nucleus of this galaxy is very bright and point-like in the near-IR,
but appears to be offset by $\sim$0.2\arcsec\ relative to the surrounding
common features in the \Iband\ image.
This could be due to large amounts of dust obscuring the
nucleus asymmetrically in the \Iband. The second brightest galaxy appears
nearly edge-on, and there is more of an obvious nuclear concentration in the
\Hband\ image than in the optical.

%%% 15 - Visit 27
\subsection{IRAS 16455+4553}

The \citet{lee94} ground-based \Rband\ image shows what appears to be two
distinct galaxies in contact, with a tidal tail emanating from one of them.
The HST images, however, indicate that this is a single object, with a pair of
large, fan-like tidal tails.
What appeared to be a second galaxy in the ground-based image is actually
just a high surface brightness tidal loop.
The two tails suggest that this must be the result of a merger event.
The WFPC2 optical image shows spiral arm structure
in the main body of the galaxy. This image also shows two peaks near the
center, one brighter than the other, suggestive of a double nucleus. The
NICMOS \Hband\ image, however, shows only a single peak, indicating that the
second bright spot in the optical image is probably a star cluster.
The rest of the object is quite featureless in the \Hband\ image, showing only
a hint of the spiral arm structure.

%%% 16 - Visit 28
\subsection{IRAS 16541+5301 (Sy 2)}

\citet{lee94} placed this system in their interaction class zero, because
a ground-based \Rband\ image shows a single body with one or more possible
tidal extensions, but no nearby companions.
The high-resolution HST images clearly reveal that this is in fact a 
compact, strongly interacting pair of galaxies that are still
separate but are in close contact. The pair is surrounded by low
surface-brightness tidal debris.
The galaxy to the south (upper right) appears to have a single
nucleus in both WFPC2 and NICMOS images.
In the \Hband\ image it is unresolved and
bright enough to see the first airy ring of the instrument diffraction pattern.
The central region of the galaxy to the north (lower left) is quite complex in
the optical \Iband\ image, showing patterns of spiral arms and/or the base of
tidal tails, bright clusters, and patchy dust lanes. In the \Iband\ image the
nucleus of this galaxy appears to be bifurcated by a narrow dust lane. 
Evidence for this dust lane is still visible in the \Hband\ image, indicating
a large optical depth. The central region of this galaxy is also much more
extended towards the north in the \Hband\ image than in the optical, again
indicating the presence of large amounts of dust that must be affecting the
optical image. The complicated spiral arm and dust lane structures seen
further out in the disk of this galaxy in the \Iband\ image are only faintly
visible in the near-IR.

%%% 17 - Visit 29
\subsection{IRAS 18580+6527 (Sy 2)}

This is a very large, strongly distorted system containing several luminous
cores.
Globally, the system is distinguished by a single large tidal tail, with
amorphous extended emission opposite the tail.
Previous ground-based \Rband\ images showed two distinct galaxy-like
objects connected by the tidal bridge to the north, with the
object to the east containing two nuclear cores \citep{lee94,aur96}.
The HST images confirm the double nuclei in the eastern object, while the
central regions of the system to the west have been revealed to contain
at least four comparably bright cores that could be the remnants of additional
progenitor galaxies.
Several of these cores show fainter, nearby hotspots.
Six out of the total of seven bright regions are equally distinct in both
\Iband\ and \Hband\ images, indicating that their appearance is most likely not
due to dust patterns.
Only one of the cores, to the south (lower right) in the \Hband\ image, has a
mottled appearance in the \Iband\ image that suggests a complex dust
arrangement.
This system is quite likely the result of collisions of at least four or
more progenitor systems.

%%% 18 - Visit 31
\subsection{IRAS 19297$-$0406}

A ground-based \Rband\ image of this system shows a crowded field, with one
main object that has chaotic morphology, but apparently a single core
\citep{mel90}.
The HST images show that this central object is a very complicated, compact
object, composed of several luminous cores embedded within a common, diffuse
envelope.
Only one obvious large tidal tail is seen in the \Hband\ image, but the
\Iband\ image show hints of other smaller tails or loops further in towards the
central region of the system.
The central region of the system is equally complex in both the I and \Hband s,
again suggesting that the multiple cores or hotspots are not simply the
result of complex dust obscuration patterns.
While the overall pattern of nuclei or cores is fairly similar in the
two different wavebands, the relative intensities are quite different.
In fact one of the most prominent nuclear remnants visible in the \Hband\ image
is almost completely obscured in the \Iband.
The brightest core appears point-like in both bands and has a short tidal
feature extending from it.

%%% 19 - Visit 32
\subsection{IRAS 20037$-$1547 (Sy 1)}

This system is dominated by the star-like nucleus of the AGN source,
with one other distinct companion galaxy located within a common halo of
tidal debris.
Several other galaxies are located nearby.
The two main progenitors are in close contact, with a large, diffuse tidal
tail wrapping around their periphery.
The non-AGN progenitor has a complex nuclear morphology in the \Iband, but
is much more regular in appearance in the \Hband\ image, which indicates the
presence of large amounts of dust.

%%% 20 - Visit 33
\subsection{IRAS 20087$-$0308 (Sy 2 or LINER?)}

This system is apparently in the advanced stages of merging, with a single
central body surrounded by diffuse, low surface-brightness emission, some
of which is in the form of tidal tails.
In ground-based optical and near-IR images the main body appears
to have an elongated or distorted nuclear region, which after the application
of resolution enhancement techniques appears to be double
\citep{mel90,duc97}.
In the WFPC2 \Iband\ image the nuclear region appears quite complicated in
appearance, showing several bright knots or condensations.
The NICMOS \Hband\ image reveals a distinct single nucleus, indicating that the
knot structure seen at shorter wavelengths is due to dust obscuration.
The nucleus appears to be surrounded by a bright ring structure, about
1\arcsec\ in diameter.
In the \Iband\ the regions surrounding the nucleus are also complicated by
dust obscuration, but in the \Hband\ image these regions are quite smooth.

%%% 21 - Visit 34
\subsection{IRAS 20100$-$4156 (LINER?)}

This system is apparently composed of at least two galaxies nearing the point
of final merging.
Ground-based images clearly show the two disk-like galaxies in near or
grazing contact, with some mild evidence for distortions or tidal debris
around the outer edges \citep{mel90,duc97}.
The HST images show that the brightest (southern) galaxy has a long tidal tail
wrapping around it and several bright knots (presumably star forming regions)
within opposing spiral arms.
The second brightest (northern) galaxy has one main nuclear region, with two
very small knots located very close to it.
Another clump of less luminous knots located a bit further away (to the
northwest) could be the remnants of yet a third galaxy.
The morphology of these regions is quite a bit more complicated in the 
\Iband\ than in the \Hband, suggesting the presence of complex dust patterns.

%%% 22 - Visit 35
\subsection{IRAS 20109$-$3003}

This is a double-tailed system, with a single distinct nucleus, thus indicating
that it is in the final stages of merging.
The central body shows subtle features in the \Hband, including two bright,
but very compact knots on either side of the main nucleus, and a pair of
opposing arms or plumes extending outwards from the nucleus.
These features show a bit more contrast in the \Iband.

%%% 23 - Visit 36
\subsection{IRAS 20176$-$4756}

This system is composed of a dominant spiral galaxy, with a severely
disrupted small companion.
The companion is nearly in contact with the main galaxy.
The companion also appears to be quite a bit fainter relative to the main
galaxy in the \Hband\ image as compared to the \Iband, suggesting that it may
be relatively blue in color.
In the \Iband\ the main galaxy shows distinct spiral arm structure, as well
as faint knots and loops within the central body.
The \Hband\ image, however, shows considerably less complicated structure,
although the main spiral arms are still visible.
Almost no structural information is contained in the \Hband\ image of the
companion galaxy, due to it's very low signal level.

%%% 24 - Visit 39
\subsection{IRAS 20414$-$1651 (LINER)}

This system has been imaged on several occasions from the ground, from
optical to near-IR wavelengths \citep{mel90, duc97,sur00}.
Most of these data show what appears to be either two overlapping galaxies,
or one main galaxy with tidal connections to or contact with one or more
fainter objects or condensations.
The HST images reveal that this very compact system is composed of a single
central body that is highly elongated, with hints of faint, amorphous
emission surrounding it in various places.
In the \Iband\ the central body has a complex morphology, containing many
knots and patches of obscuration, and no clear sign of any dominant remnant
nuclei.
The morphology in the \Hband, on the other hand, is quite regular, showing a
distinct central core.
The central core appears to be split, however, into two components with a
separation of only $\sim$0.15--0.20\arcsec.

%%% 25 - Visit 44
\subsection{IRAS 22206$-$2715}

This is a complex interacting pair of galaxies.
Initial ground-based imaging showed a single dominant object, with multiple
tidal extensions, all of which appeared featureless \citep{cle96b}.
More recent higher-resolution data revealed that there are actually two
remant galaxies present, with the northern object having a roughly
spherical or elliptical morphology, and the southern appearing very elongated,
suggesting an edge-on disk \citep{sur00}.
The HST images show that the northern, relatively face-on galaxy, has several
clouds of diffuse tidal debris on either side of the main body.
The southern edge-on galaxy has a long tidal tail extending from the its
central body.
The majority of this tail is quite straight and narrow, but has a remarkably
sharp turn or twist in direction near the point of contact with the central
body of the galaxy.
The morphology in the WFPC2 \Iband\ is very complex, with evidence for many
star-forming knots and dust lanes throughout the central bodies of both
galaxies, as well as the tidal structures.
The edge-on galaxy has so much dust obscuration in the \Iband\ that it is
difficult to determine the existence or position of the remnant nucleus.
In the \Hband\ the effects due to dust are drastically reduced, allowing
the remnant core of this galaxy to be much more easily distinguished.
Several knots and hot-spots do still appear in the \Hband\ image, however,
and are coincident with the most luminous knots seen in the \Iband\ image.
One of the largest of these regions is located approximately midway between
the two progenitor galaxies, suggesting a region of shock-induced star
formation at the point of contact.

%%% 26 - Visit 46
\subsection{IRAS 23128$-$5919 (Sy 2?)}

This is a pair of strongly interacting galaxies very similar in appearance
to the nearby NGC~4038/4039 (``Antennae'') pair.
The two progenitor galaxies appear to be of similar mass and size, and each
has a very long tidal tail extending over several galaxy diameters.
This system has been the target of several ground-based imaging studies.
Broad-band data have shown the two distinct progenitor galaxies, with a
nuclear separation of $\sim$4.5\arcsec\ \citep{zen93,duc97}.
These data indicate that the southern galaxy is much redder in color
than the northern galaxy.
\Ha\ maps show that the majority of emission is concentrated in the two nuclei,
although some emission also appears between them \citep{mih98}.

In the HST images the central bodies of the two galaxies have a very complex
morphology in both the I and \Hband, containing many bright knots of emission
(super star clusters?), as well as complex dust obscuration patterns.
In spite of the complicated morphology the two progenitor nuclei are
relatively distinct and straightforward to identify.
The nucleus of the southern (lower right) galaxy is
relatively bright and point-like, indicating the possible existence of an
active nucleus.
There are numerous emission knots located very near the nuclei, as well as
throughout the central bodies and on their periphery.
The coincidence of many of these knots in both the I and \Hband\ images
suggests that they are not simply due to local regions of low obscuration,
but are in fact bright star clusters.
This is supported by the similar morphologies seen in ground-based I and
K images \citep{zen93}, suggesting relatively little dust content.

\citet{mih98} found that \Ha\ emission line profiles in the southern galaxy
of this pair were very complex, showing asymmetric shapes or double-line
profiles. In several regions, two distinct kinematic profiles were indicated
by the data. They also found that the \Ha\ emission associated with the
southern galaxy nucleus is spatially extended in the east-west direction.
These observed effects are quite likely a result of the existence of
the multiple subcondensations and apparent star forming regions as seen in the
HST images.

%%% 27 - Visit 49
\subsection{IRAS 23230$-$6926 (LINER?)}

This is a compact system composed of a single, amorphous body, with faint
loops and shells of tidal material surrounding.
This morphology suggests that it is at an advanced merger stage.
Previous ground-based images showed what appeared to be two galaxies in close
contact, or a single object with an elongated, but unresolvable, nuclear region
\citep{mel90,duc97}.
The WFPC2 \Iband\ image shows only one distinct and dominant nuclear core, but
the near-nuclear morphology is quite complicated by dust obscuration and
bright emission knots.
A secondary nucleus could be obscured.
The NICMOS \Hband\ image confirms this suspicion, revealing what appears to be
a second remnant core located $\sim$0.6\arcsec\ from the first.
Both nuclei appear to be somewhat elongated along the direction of
their separation.
Note, however, that there is also the appearance of a dust lane running
between the two nuclei, and therefore the two bright regions could be a single
nucleus that is bisected by the dust lane.
Some of the bright knots and regions located further out in the central body
in the \Iband\ image are also visible in the NICMOS near-IR image.

\section{Summary}

An HST NICMOS \Hband\ survey of 27 ultraluminous infrared galaxies has
shown the following.

\begin{itemize}
\item{}
At least 85\%\ of the systems are obviously composed of two or more
interacting galaxies, with as many as 93\%\ showing some
signs of interaction history.

\item{}
Approximately 30\%\ of the systems have existing spectra with characteristics
of Seyfert 1 or 2 type nuclei.
An additional 7\%\ of the systems show a bright, unresolved nuclear source
in the HST \Hband\ images, yielding a fraction of 37\%\ of these
systems with evidence for an AGN.
These data do not rule out, however, the possible existence of heavily
dust-enshrouded AGN or QSO.

\item{}
There is often a stark contrast in the optical and near-IR morphologies of the
ULIRGs.
HST \Iband\ morphologies of $\sim$75--80\%\ these systems are quite
complicated and often chaotic, while only $\sim$25--30\%\ of the
\Hband\ images show similar features.
The relative lack of small-scale features in the \Hband\ images suggests
that a majority of such optical features are due to complicated dust
morphologies and a widely distributed ensemble of hot, young star clusters.

\item{}
The reduced sensitivity of \Hband\ images to the effects of dust
obscuration as well as the blue colors of young star clusters has revealed
more of the true nature of the underlying global stellar mass distributions in
these systems. Specifically, three systems that were previously classified as
containing single nuclei have been shown to contain the remnant cores of two
or more galaxies. A total of seven systems that have the global morphology
of a single body have been shown to contain double or multiple nuclei.
Furthermore, in about a dozen systems with complicated
optical morphologies and therefore no clear identification or segregation of
remnant nuclei and young super star clusters, the \Hband\ images have clearly
revealed the dominant core nuclei.

\item{}
The difference in \Iband\ and \Hband\ morphologies of these systems serves as
yet another reminder that galaxy morphology is quite wavelength-dependent, and
therefore any implications for galaxy structure or evolution
based on the optical morphologies of high-redshift galaxies must be treated
with care \citep[e.g.][]{gia96}.
The complicated optical morphologies taken by themselves might lead one to
believe that many of the systems are in a severely disturbed or disrupted state.
The \Hband\ images, on the other hand, reveal that many systems
have underlying mass distributions more typical of ordinary disk-type galaxies
(albeit mass distributions that have suffered the tidal warping effects
of a close encounter with a companion).

\end{itemize}

\acknowledgements

We wish to thank Molly Shea-Martinez (Program Coordinator) and Anatoly
Suchkov (Contact Scientist) for their support of this observing program at
the STScI.
KB thanks Ron Allen and the Space Telescope Science Institute for their
hospitality in sponsoring his sabbatical visit.
This research was supported in part by NASA through grant number
GO-07896.02-96A from the STScI.
This research has made use of the NASA Astrophysics Data System Abstract
Service and the NASA/IPAC Extragalactic Database (NED), which is operated by
the Jet Propulsion Laboratory, CalTech, under contract with the National
Aeronautics and Space Administration.

\clearpage

\begin{figure}
\epsscale{0.75}
\plotone{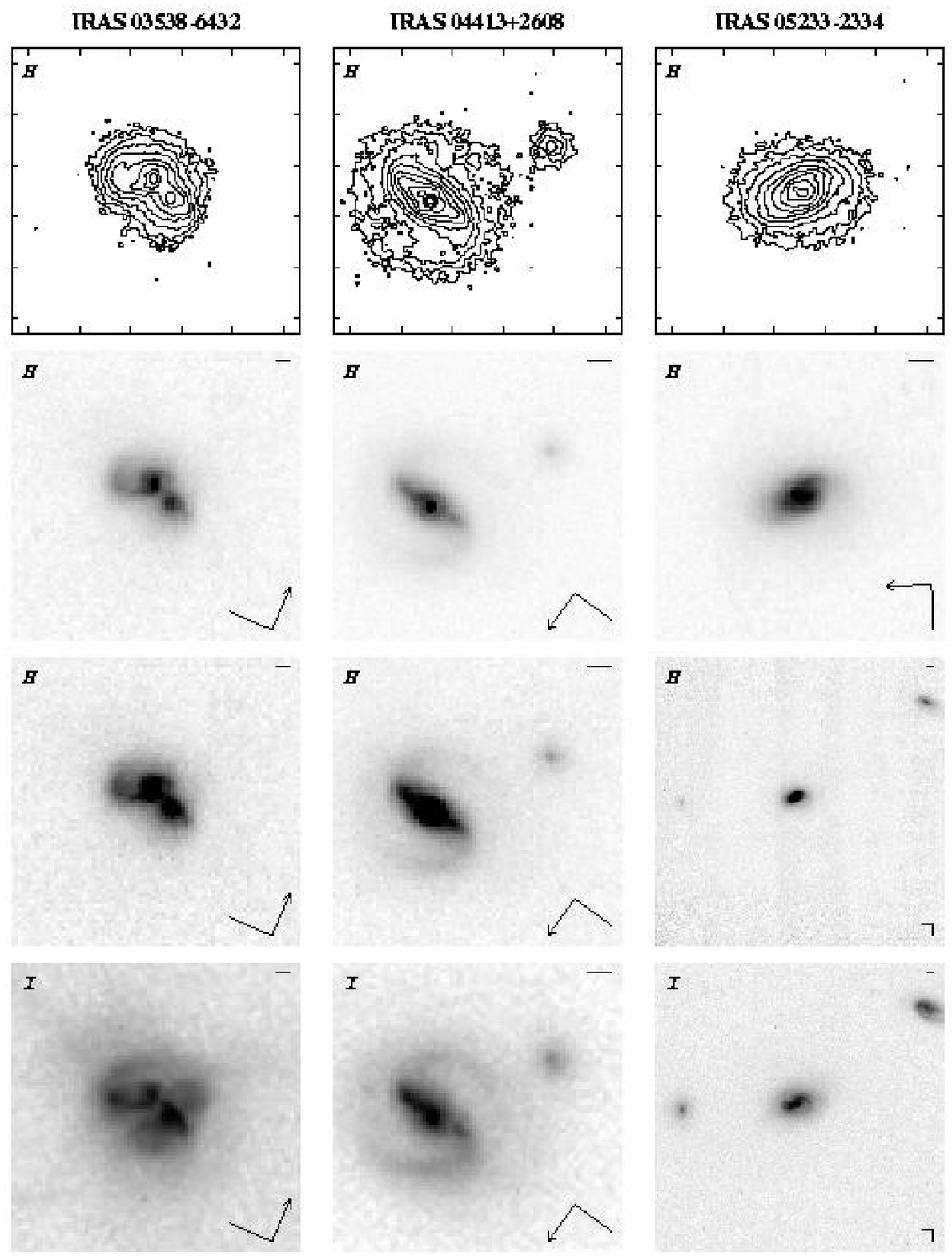}
\caption{NICMOS and WFPC2 images of the ULIRGs.
The four panels for each system show, from top to bottom, a NICMOS
\Hband\ contour plot, \Hband\ shallow and deep images,
and a WFPC2 \Iband\ image.
The contour levels start at a surface brightness of
$m_H=18.72$ mag arcsec$^{-2}$ and increase at 0.5 mag intervals.
The contour axis tic marks are at 1\arcsec\ intervals.
The compass arrows indicate North and East, and are 1\arcsec\ long.
The physical scale bar in the upper right of each image is 2 kpc in length.
\label{fig-images}}
\end{figure}

\begin{figure}
%\epsscale{0.9}
\plotone{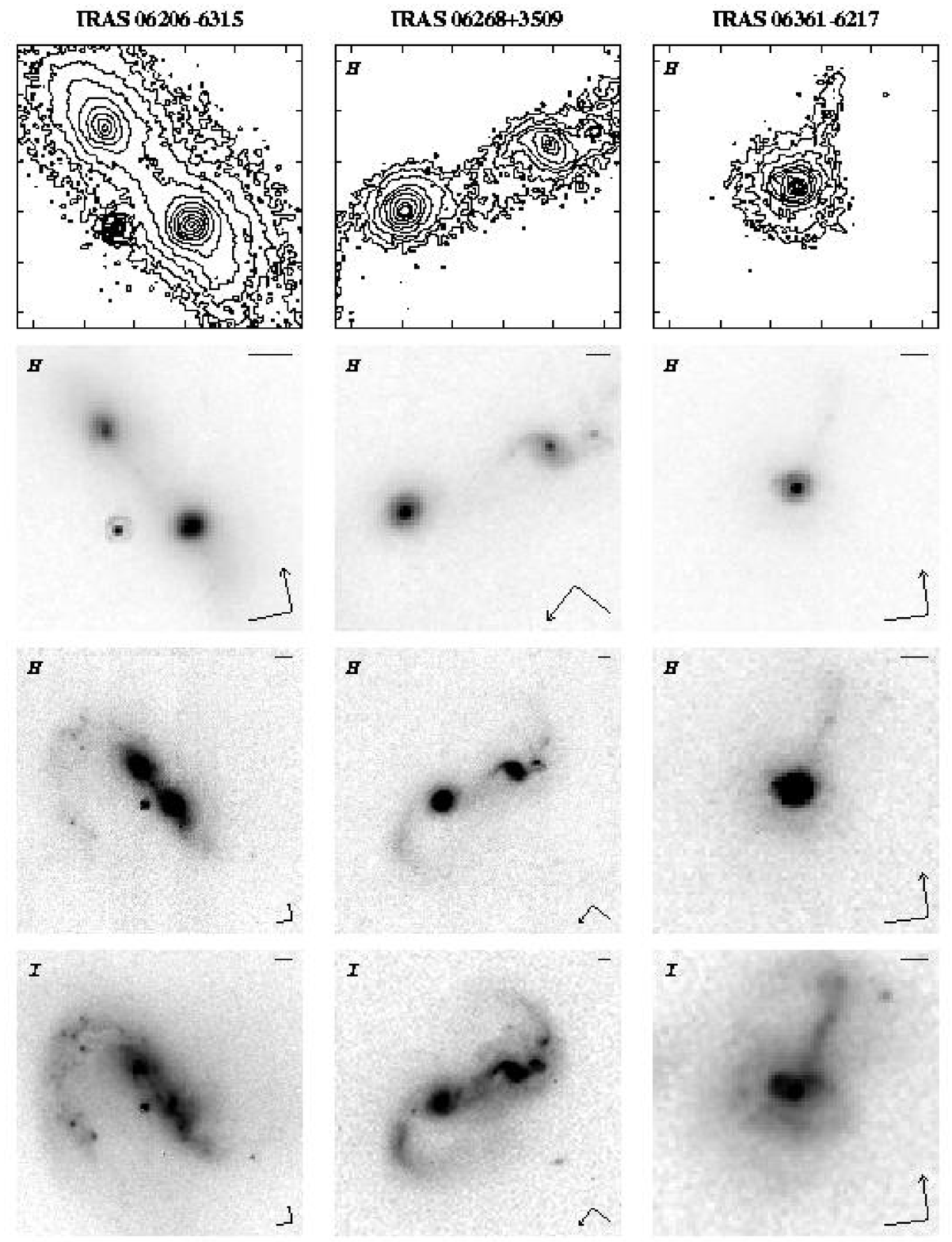}
\setcounter{figure}{0}
\caption{(continued)}
\end{figure}

\begin{figure}
%\epsscale{0.9}
\plotone{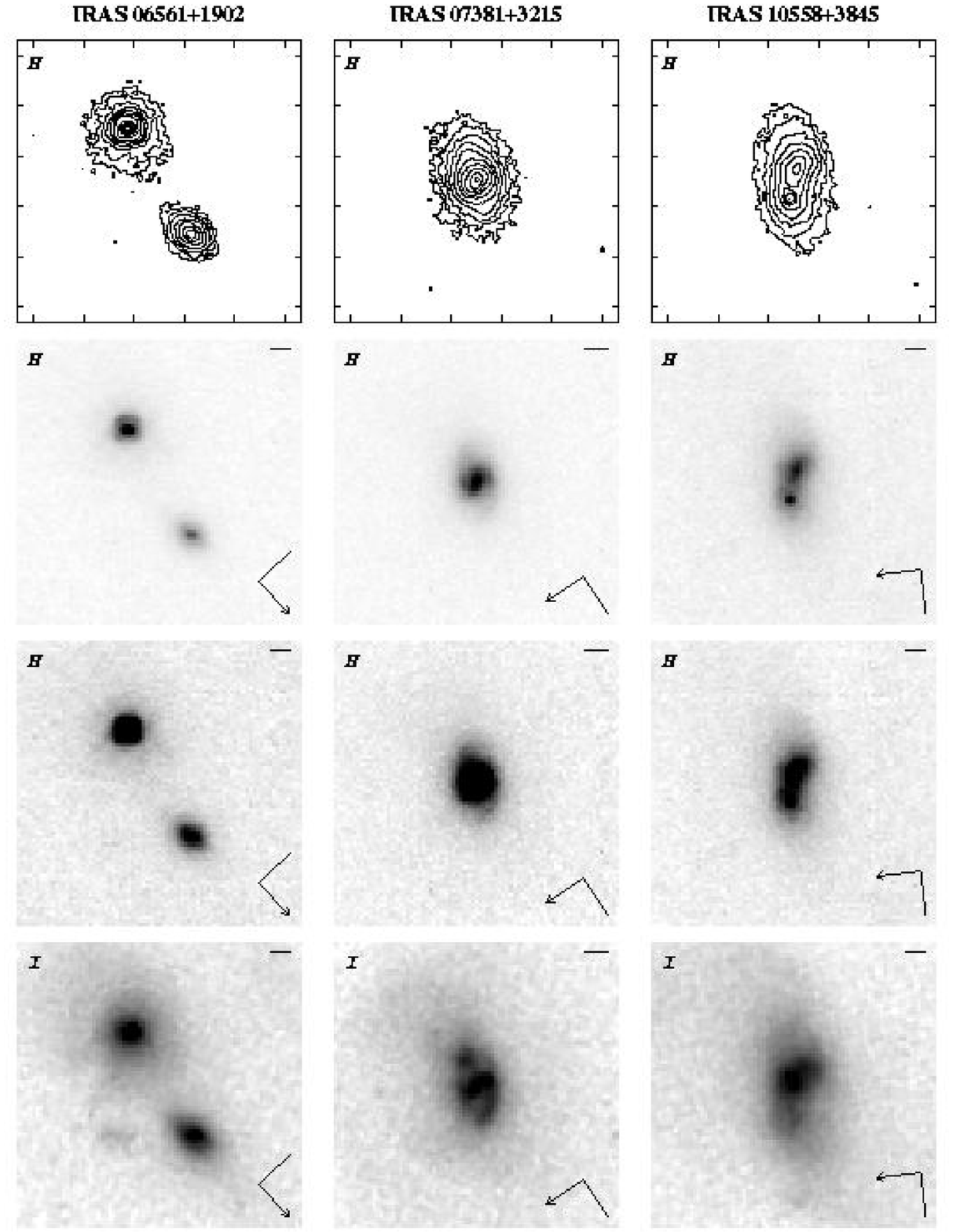}
\setcounter{figure}{0}
\caption{(continued)}
\end{figure}

\begin{figure}
%\epsscale{0.9}
\plotone{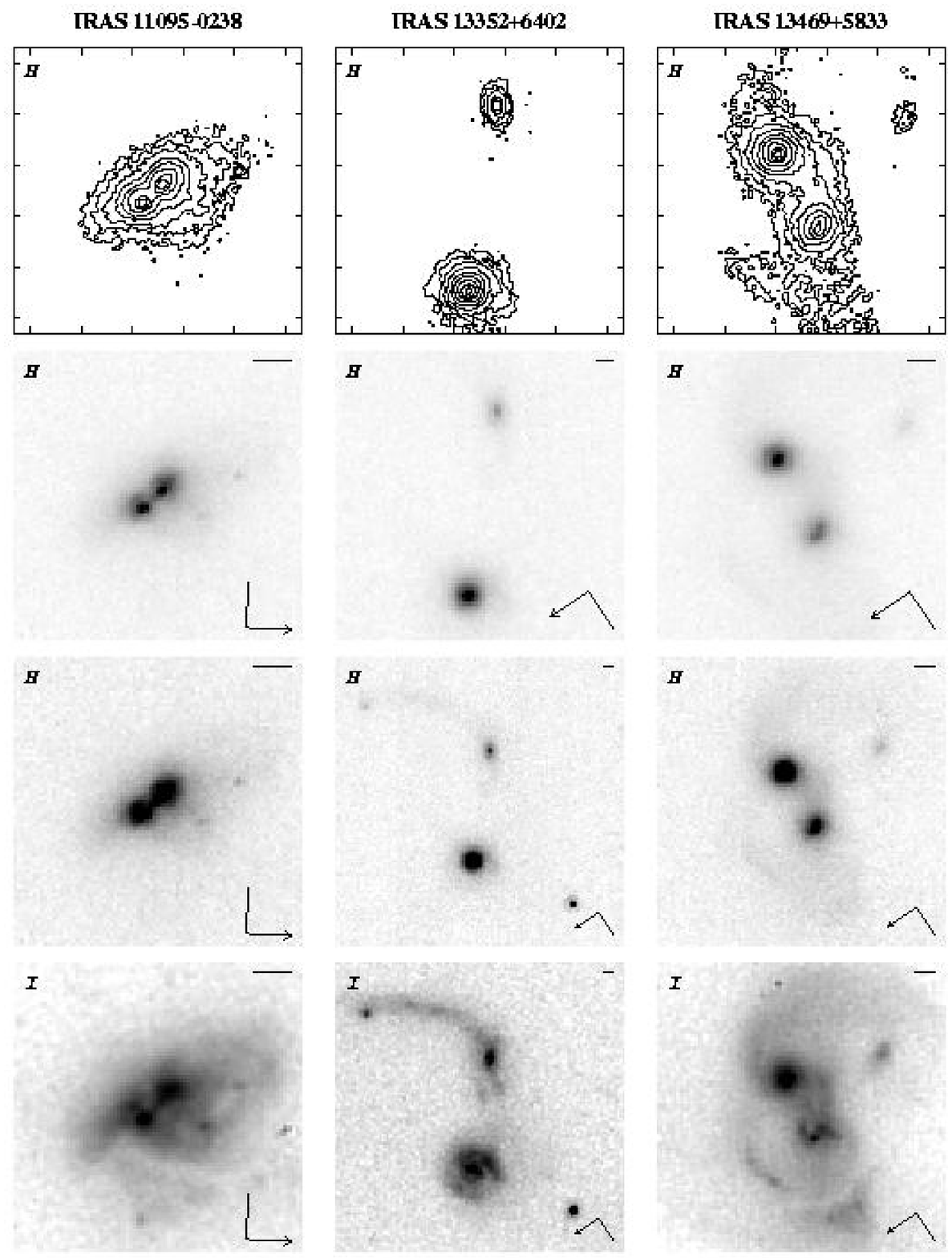}
\setcounter{figure}{0}
\caption{(continued)}
\end{figure}

\begin{figure}
%\epsscale{0.9}
\plotone{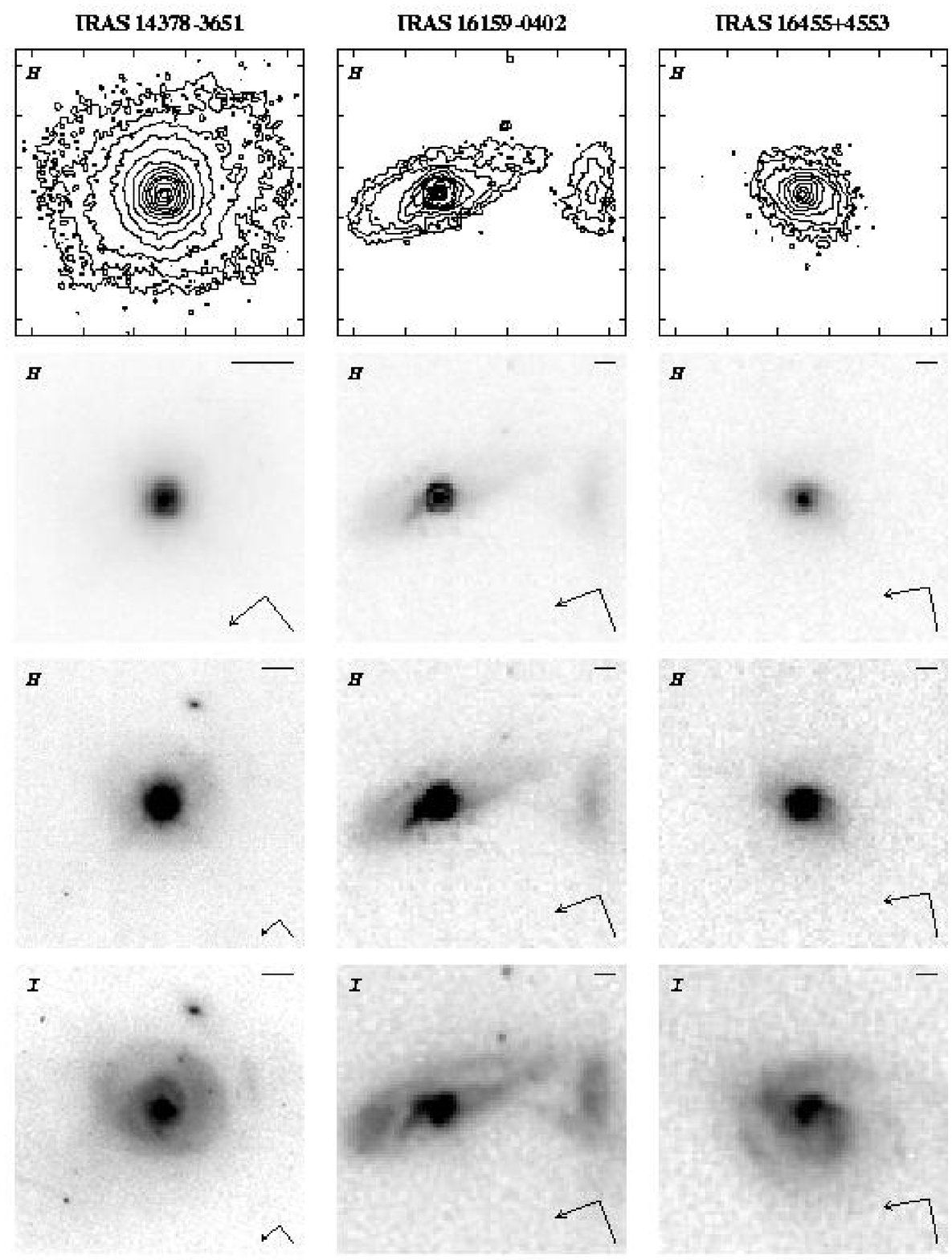}
\setcounter{figure}{0}
\caption{(continued)}
\end{figure}

\begin{figure}
%\epsscale{0.9}
\plotone{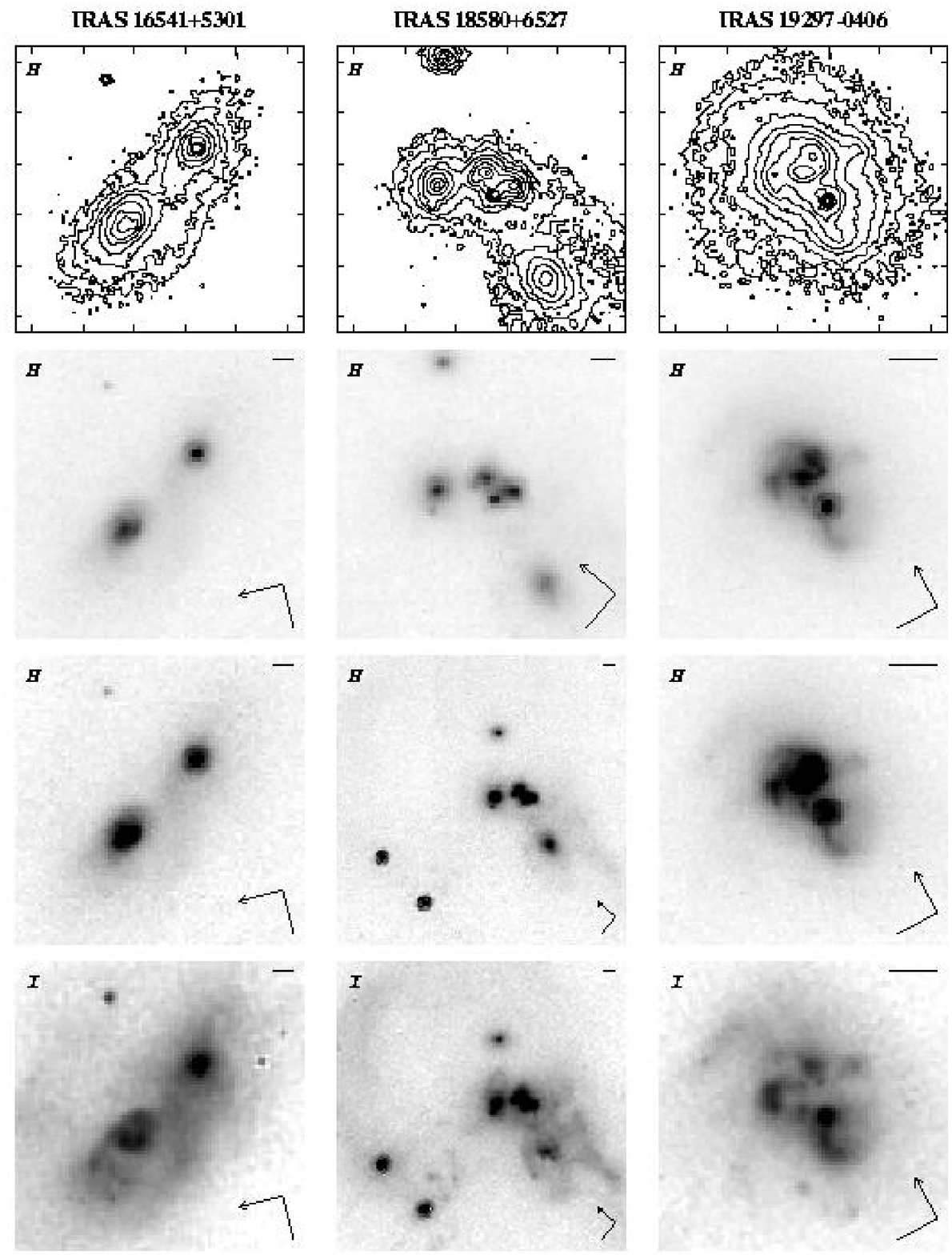}
\setcounter{figure}{0}
\caption{(continued)}
\end{figure}

\begin{figure}
%\epsscale{0.9}
\plotone{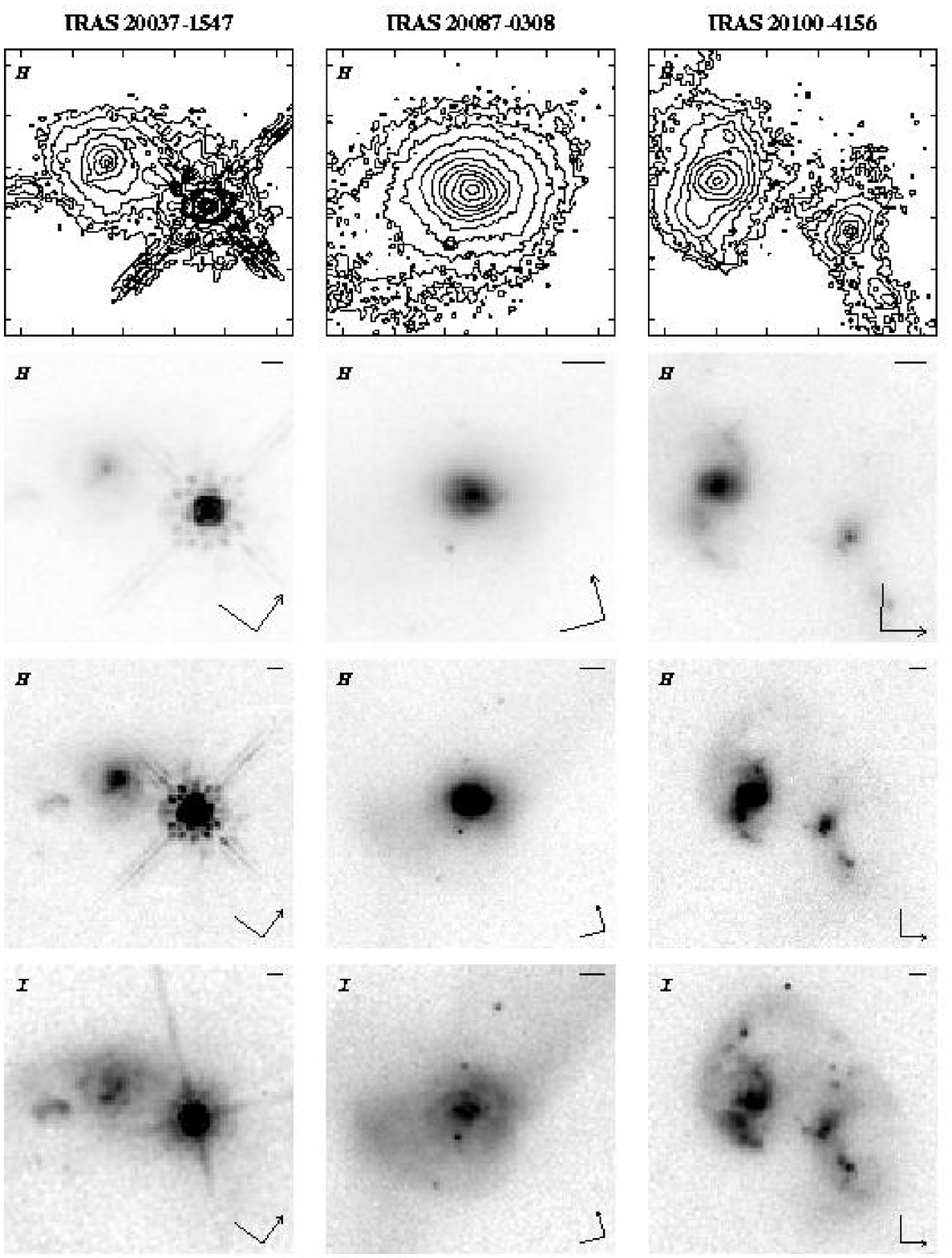}
\setcounter{figure}{0}
\caption{(continued)}
\end{figure}

\begin{figure}
%\epsscale{0.9}
\plotone{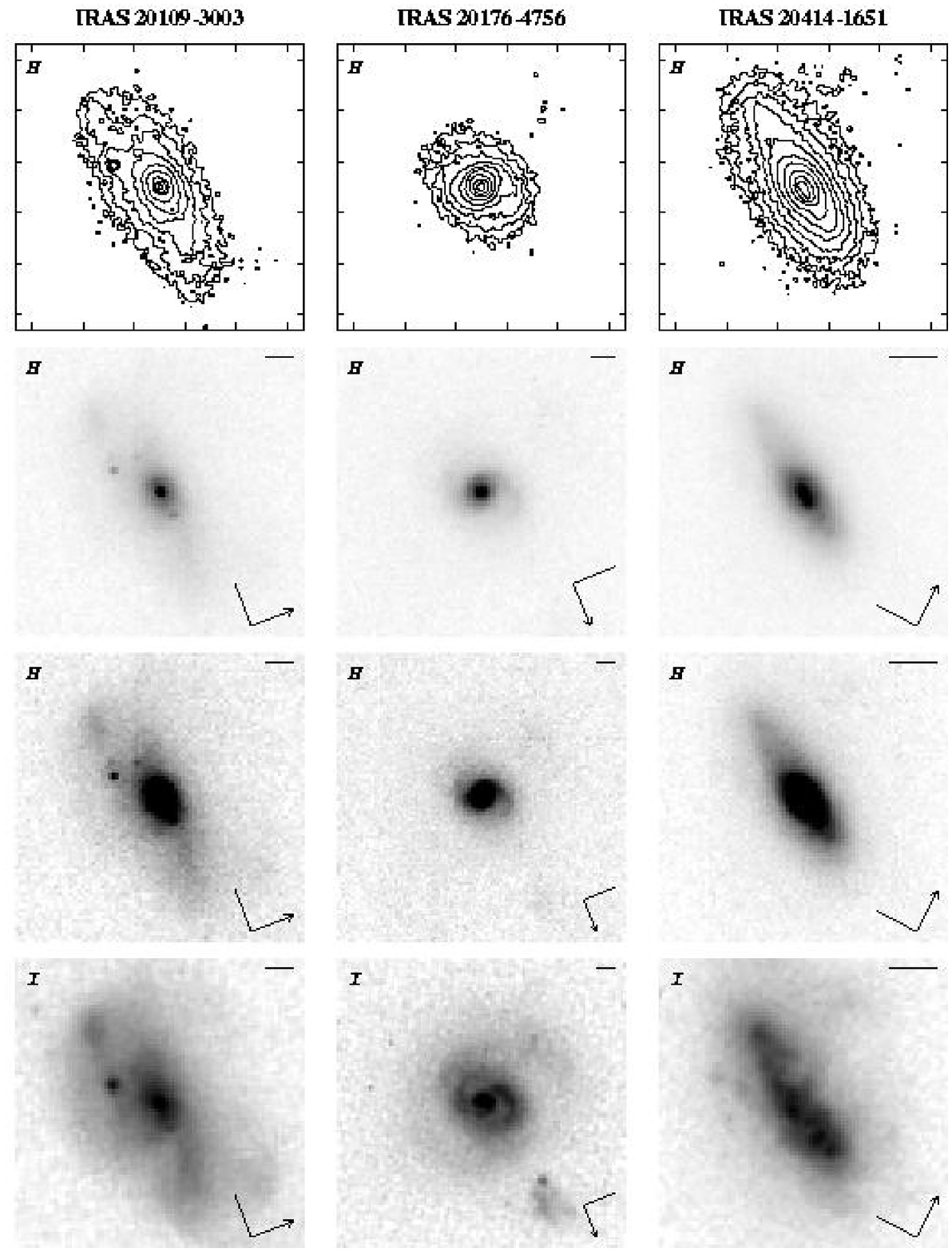}
\setcounter{figure}{0}
\caption{(continued)}
\end{figure}

\begin{figure}
%\epsscale{0.9}
\plotone{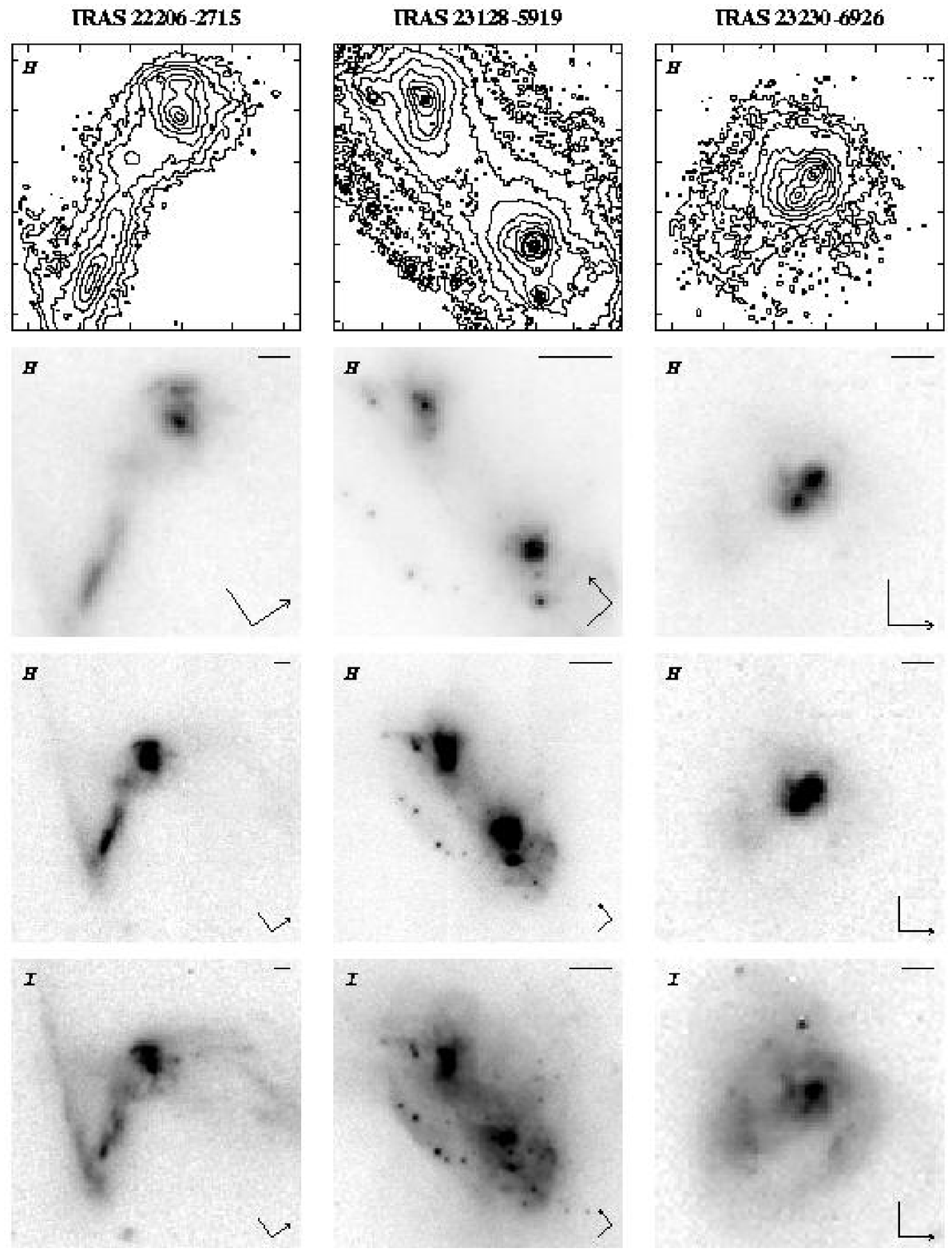}
\setcounter{figure}{0}
\caption{(continued)}
\end{figure}

\clearpage

\begin{deluxetable}{lcccccccccc}
\tabletypesize{\scriptsize}
%\tablenum{1}
\tablecolumns{11}
\tablecaption{ULIRG Sample Data. \label{tbl1}}
\tablewidth{460pt}
\tablehead{
\colhead{\ } & \colhead{\ } & \colhead{\ftwelve} & \colhead{\ftwenty} &
\colhead{\fsixty} & \colhead{\fhundred} & \colhead{\Lsixty\tablenotemark{a}} &
\colhead{\ } & \colhead{Spectral} & \colhead{\ } & \colhead{\ } \\
\colhead{Name} & \colhead{$z$} & \colhead{(Jy)} & \colhead{(Jy)} & \colhead{(Jy)} &
\colhead{(Jy)} & \colhead{(log \Lsun)} & \colhead{\ftwenty/\fsixty} &
\colhead{Class\tablenotemark{b}} & \colhead{Ref.} & \colhead{$M_H$}
}
\startdata
IRAS 03538$-$6432& 0.301& $<$0.06& \pl0.06& 0.99& 1.30& 12.46&\pl0.07 & \nod &\nod &$-$25.33\\
IRAS 04413$+$2608& 0.171& $<$0.31& $<$0.31& 0.82&$<$2.23\pl&11.87&$<$0.38 & S2 & 1 &$-$24.77\\
IRAS 05233$-$2334& 0.172& $<$0.06& $<$0.06& 0.53& 1.16& 11.69&$<$0.11 & \nod &\nod &$-$24.37\\
IRAS 06206$-$6315& 0.092& $<$0.07& \pl0.29& 3.96& 4.58& 12.00&\pl0.07 & S2   &   2 &$-$24.17\\
IRAS 06268$+$3509& 0.170& $<$0.26& $<$0.25& 0.94& 1.09& 11.92&$<$0.27 & \nod &\nod &$-$24.47\\
IRAS 06361$-$6217& 0.160& $<$0.11& \pl0.17& 1.66& 2.01& 12.11&\pl0.10 & \nod &\nod &$-$23.89\\
IRAS 06561$+$1902& 0.188& $<$0.29& $<$0.28& 1.01& 1.32& 12.05&$<$0.28 & \nod &\nod &$-$24.08\\
IRAS 07381$+$3215& 0.170& $<$0.07& $<$0.14& 0.67& 0.83& 11.78&$<$0.21 & \nod &\nod &$-$24.14\\
IRAS 10558$+$3845& 0.207& $<$0.05& $<$0.09& 0.62& 0.75& 11.92&$<$0.14 & \nod &\nod &$-$24.31\\
IRAS 11095$-$0238& 0.107& $<$0.13& \pl0.42& 3.25& 2.53& 12.04&\pl0.13 & LI   &   2 &$-$23.00\\
IRAS 13352$+$6402& 0.237& $<$0.06& $<$0.08& 0.99& 1.43& 12.24&$<$0.08 & \nod &\nod &$-$23.91\\
IRAS 13469$+$5833& 0.158& $<$0.05& $<$0.07& 1.27& 1.73& 11.99&$<$0.06 & SB   &   3 &$-$24.27\\
IRAS 14378$-$3651& 0.068& $<$0.12& \pl0.52& 6.19& 6.34& 11.92&\pl0.08 & S2   &   2 &$-$23.66\\
IRAS 16159$-$0402& 0.213& \pl0.18& \pl0.30& 0.98&$<$1.77\pl&12.14&\pl0.31 & \nod &\nod &$-$24.86\\
IRAS 16455$+$4553& 0.191& $<$0.06& \pl0.08& 0.94& 1.59& 12.03&\pl0.09 & \nod &\nod &$-$23.70\\
IRAS 16541$+$5301& 0.194& $<$0.05& $<$0.07& 0.68& 1.28& 11.90&$<$0.11 & S2   &   1 &$-$24.72\\
IRAS 18580$+$6527& 0.176& $<$0.05& \pl0.07& 0.76& 1.40& 11.87&\pl0.09 & S2,SB&   1 &$-$25.12\\
IRAS 19297$-$0406& 0.086& $<$0.13& \pl0.59& 7.05& 7.72& 12.19&\pl0.08 & SB   &   4 &$-$24.09\\
IRAS 20037$-$1547& 0.192& $<$0.13& $<$0.28& 1.65& 1.98& 12.28&$<$0.17 & S1   &   1 &$-$26.28\\
IRAS 20087$-$0308& 0.106& $<$0.13& \pl0.24& 4.70& 6.54& 12.20&\pl0.05 & S2?LI&   2 &$-$24.51\\
IRAS 20100$-$4156& 0.130& $<$0.13& \pl0.34& 5.23& 5.16& 12.42&\pl0.07 & SB$>$LI& 2 &$-$24.30\\
IRAS 20109$-$3003& 0.143& $<$0.06& $<$0.24& 0.85&$<$1.33\pl&11.72&$<$0.28 & \nod & \nod&$-$23.74\\
IRAS 20176$-$4756& 0.178& $<$0.07& $<$0.08& 0.88& 1.47& 11.93&$<$0.09 & \nod & \nod&$-$23.98\\
IRAS 20414$-$1651& 0.087& $<$0.65& \pl0.35& 4.36& 5.25& 11.99&\pl0.08 & LI$>$SB& 2 &$-$23.51\\
IRAS 22206$-$2715& 0.131& $<$0.10& $<$0.16& 1.75& 2.33& 11.96&$<$0.09 & SB   &   3 &$-$24.27\\
IRAS 23128$-$5919& 0.045& \pl0.24& \pl1.59&10.80\phm{1}&10.99\phm{1}&11.80&\pl0.15&SB+LI+S2& 2 &$-$23.81\\
IRAS 23230$-$6926& 0.106& $<$0.06& \pl0.29& 3.74& 3.42& 12.10&\pl0.08 & LI+SB&   2 &$-$23.80\\
\enddata
\tablenotetext{a}
{$\Lsixty = 1.14 \times 10^{14} \times (1-(1+z)^{-0.5})^{2} \times (1+z)^{2} \times \fsixty$,
assuming $\Ho = 70~km~s^{-1}~Mpc^{-1}$ and $q_0 = 0.5$.}
\tablenotetext{b}
{Key: SB = Starburst; LI = LINER; S1 = Seyfert 1; S2 = Seyfert 2.
{\it a,b} - data for individual galaxies;
{\it a?b} - data inconclusive for distinguishing between types;
{\it a$>$b} - data show type {\it a} dominant over type {\it b};
{\it a$+$b} - data show characteristics of both types.}
\tablerefs{
(1) Lawrence et al. 1999;
(2) Duc, Mirabel, \& Maza 1997;
(3) Veilleux, Kim, \& Sanders 1999;
(4) Veilleux et al. 1995.}
\end{deluxetable}

\clearpage

\begin{deluxetable}{lllc}
\tabletypesize{\scriptsize}
\tablecolumns{4}
\tablecaption{Morphological Descriptions. \label{tbl2}}
\tablewidth{315pt}
\tablehead{
\colhead{\ } & \colhead{\ } & \colhead{\ } & \colhead{Unresolved} \\
\colhead{Name} & \colhead{Global} & \colhead{Nuclear\tablenotemark{a}} &
\colhead{Nucleus?}
}
\startdata
IRAS 03538$-$6432& Single main body with tail     & Double    & no  \\
IRAS 04413$+$2608& Pair                           & Single    & yes \\
IRAS 05233$-$2334& Single in group                & Single    & no  \\
IRAS 06206$-$6315& Pair with tails                & Single    & no  \\
IRAS 06268$+$3509& Pair with tails                & Single    & no  \\
IRAS 06361$-$6217& Single with possible companion & Single    & no  \\
IRAS 06561$+$1902& Pair                           & Single    & no  \\
IRAS 07381$+$3215& Single                         & Single    & no  \\
IRAS 10558$+$3845& Single main body               & Double    & no  \\
IRAS 11095$-$0238& Single main body with tails    & Double    & no  \\
IRAS 13352$+$6402& Pair with tail                 & Single    & no  \\
IRAS 13469$+$5833& Pair with tails                & Single    & no  \\
IRAS 14378$-$3651& Single                         & Single    & yes \\
IRAS 16159$-$0402& Pair                           & Single    & yes \\
IRAS 16455$+$4553& Single with tails              & Single    & no  \\
IRAS 16541$+$5301& Pair                           & Single    & yes \\
IRAS 18580$+$6527& Group with tails               & Multiple? & no  \\
IRAS 19297$-$0406& Single main body               & Double    & yes \\
IRAS 20037$-$1547& Pair?                          & Single    & yes \\
IRAS 20087$-$0308& Single main body with tails    & Single    & no  \\
IRAS 20100$-$4156& Pair with tails                & Single    & no  \\
IRAS 20109$-$3003& Single with tails              & Single    & no  \\
IRAS 20176$-$4756& Single with possible companion & Single    & no  \\
IRAS 20414$-$1651& Single with tidal debris       & Double?   & no  \\
IRAS 22206$-$2715& Pair with tails                & Single    & no  \\
IRAS 23128$-$5919& Pair with tails                & Single    & yes \\
IRAS 23230$-$6926& Single main body with tails    & Double    & no  \\
\enddata
\tablenotetext{a}
{For systems designated as a pair, a nuclear morphology of ``Single''
indicates that each galaxy has a single nucleus.}
\end{deluxetable}


\clearpage

\begin{thebibliography}{DUM}

\bibitem[Ashby, Houck, \& Matthews(1995)]{ash95}
Ashby, M., Houck, J. R., \& Matthews, K.
1995, \apj, 447, 545

\bibitem[Auriere et~al.(1996)]{aur96}
Auriere, M., Hecquet, J., Coupinot, G., Arthaud, R., \& Mirabel, I. F.
1996, \aap, 312, 387

\bibitem[Barger et~al.(1998)]{bar98}
Barger, A. J., Cowie, L. L., Sanders, D. B., Fulton, E., Taniguchi, Y.,
Sato, Y., Kawara, K., \& Okuda, H.
1998, Nature, 394, 248

\bibitem[Barger, Cowie, \& Sanders(1999)]{bar99}
Barger, A. J., Cowie, L. L., \& Sanders, D. B.
1999, \apj, 518, L5

\bibitem[Blain et~al.(1999)]{bla99}
Blain, A. W., Smail, I., Ivison, R. J., \& Kneib, J.-P.
1999, \mnras, 302, 632

\bibitem[Borne et~al.(2000)]{bor00}
Borne, K. D., Bushouse, H., Lucas, R. A., \& Colina, L.
2000, \apjl, 529, L77

\bibitem[Borne et~al.(2001)]{bor01}
Borne, K. D., Bushouse, H. A., Colina, L., \& Lucas, R. A.
2001, \apjs, in preparation

\bibitem[Boyce et~al.(1996)]{boy96}
Boyce, P. J., et al.
1996, \apj, 473, 760

\bibitem[Bushouse(1997)]{bus97}
Bushouse, H.
1997, in The 1997 HST Calibration Workshop, ed. S. Casertano, R. Jedrzejewski,
C. D. Keyes, \& M. Stevens (Baltimore: STScI), 223

\bibitem[Bushouse, Dickinson, \& van der Marel(2000)]{bus00}
Bushouse, H., Dickinson, M., \& van der Marel, R. P.
2000, in ASP Conf. Ser. 216, Astronomical Data Analysis Software and 
Systems IX, ed. N. Manset, C. Veillet, \& D. Crabtree (San Francisco: ASP), 531

\bibitem[Clements et~al.(1996a)]{cle96a}
Clements, D. L.,Sutherland, W. J., Saunders, W., Efstathiou, G. P.,
McMahon, R. G., Maddox, S., Lawrence, A., \& Rowan-Robinson, M.
1996, \mnras, 279, 459

\bibitem[Clements et~al.(1996b)]{cle96b}
Clements, D. L., Sutherland, W. J., McMahon, R. G., \& Saunders, W.
1996, \mnras, 279, 477

\bibitem[Colina et~al.(2001)]{col01}
Colina, L., et al.
2001, \apj, in press

\bibitem[Duc, Mirabel, \& Maza(1997)]{duc97}
Duc, P.-A., Mirabel, I. F., \& Maza, J.
1997, \aaps, 124, 533

\bibitem[Eales et~al.(1999)]{eal99}
Eales, S., Lilly, S., Gear, W., Dunne, L., Bond, J. R., Hammer, F.,
Le Fevre, O., \& Crampton, D.
1999, \apj, 515, 518

\bibitem[Farrah et~al.(2001)]{far01}
Farrah, D., et al.
2001, \mnras, in press

\bibitem[Fullmer \& Lonsdale(1989)]{ful89} Fullmer, L., \& Lonsdale, C.
1989, Cataloged Galaxies and Quasars Observed in the \IRAS\ Survey,
Version 2 (Pasadena:JPL)

\bibitem[Genzel et~al.(1998)]{gen98}
Genzel, R., et al.
1998, \apj, 498, 579

\bibitem[Giavalisco et~al.(1996)]{gia96}
Giavalisco, M., Livio, M., Bohlin, R. C., Macchetto, F. D., \& Stecher, T. P.
1996, \aj, 112, 369

\bibitem[Holland et~al.(1999)]{hol99}
Holland, W. S., et al.
1999, \mnras, 303, 659

\bibitem[Hughes et~al.(1998)]{hug98}
Hughes, D. H., et al.
1998, Nature, 394, 241

\bibitem[Kim(1995)]{kim95}
Kim, D.-C. 1995, Ph.D. thesis, U. Hawaii

\bibitem[Kim et~al.(1995)]{kimetal95}
Kim, D.-C., Sanders, D. B., Veilleux, S., Mazzarella, J. M., \& Soifer, B. T. 
1995, \apjs, 98, 129

\bibitem[Kim, Veilleux, \& Sanders(1998)]{kim98}
Kim, D.-C., Veilleux, S., \& Sanders, D. B.
1998, \apj, 508, 627

\bibitem[Klaas(1989)]{kla89}
Klaas, U.
1989, \aaps, 157, 245

\bibitem[Klass \& Els\"asser(1993)]{kla93}
Klaas, U., \& Els\"asser, H.
1993, \aaps, 99, 71

\bibitem[Lawrence et~al.(1989)]{law89}
Lawrence, A., Rowan-Robinson, M., Leech, K. J., Jones, D. H. P., \& Wall, J. V.
1989, \mnras, 240, 329

\bibitem[Lawrence et~al.(1999)]{law99}
Lawrence, A. et al.
1999, \mnras, 308, 897

\bibitem[Leech et~al.(1989)]{lee89}
Leech, K. J., Penston, M. V., Terlevich, R. J., Lawrence, A.,
Rowan-Robinson, M., \& Crawford, J.
1988, \mnras, 240, 349

\bibitem[Leech et~al.(1994)]{lee94}
Leech, K. J., Rowan-Robinson, M., Lawrence, A., \& Hughes, J. D.
1994, \mnras, 267, 253

\bibitem[Lutz et~al.(1998)]{lut98}
Lutz, D., Spoon, H. W. W., Rigopoulou, D., Moorwood, A. F. M., \& Genzel, R.
1998, \apj, 505, L103

\bibitem[Lutz, Veilleux, \& Genzel(1999)]{lut99}
Lutz, D., Veilleux, S., \& Genzel, R.
1999, \apj, 517, L13

\bibitem[Melnick \& Mirabel(1990)]{mel90}
Melnick, J., \& Mirabel, F. 
1990, \aap, 231, L19

\bibitem[Mihos \& Bothun(1998)]{mih98}
Mihos, J. C., \& Bothun, G. D.
1998, \apj, 500, 619

\bibitem[Moshir et~al.(1990)]{mos90}
Moshir, M. et al.
1990, Infrared Astronomical Satellite Catalogs, The Faint Source Catalog,
version 2 (Pasadena:JPL)

\bibitem[Murphy et~al.(1996)]{mur96}
Murphy, T. W., Jr., Armus, L., Matthews, K., Sofier, B. T., Mazarrella, J. M.,
Shupe, D. L., Strauss, M. A., \& Neugebauer, G. 1996, \aj, 111, 1025

\bibitem[Rigopoulou et~al.(1999)]{rig99}
Rigopoulou, D., Spoon, H. W. W., Genzel, R., Lutz, D., Moorwood, A. F. M.,
\& Tran, Q. D. 1999, \aj, 118, 2625

\bibitem[Rowan-Robinson(1991)]{row91}
Rowan-Robinson, M.
1991, in Dynamics of Molecular Cloud Distributions,
eds. F. Combes \& F. Fasoli, 211 (Dordrecht: Kluwer)

\bibitem[Sanders(2000)]{san00}
Sanders, D. B.
2000, in Ultraluminous Galaxies: Monsters or Babies?, in press

\bibitem[Sanders \& Mirabel(1996)]{san96}
Sanders, D. B., \& Mirabel, I. F. 1996, \araa, 34, 749

\bibitem[Sanders et~al.(1988a)]{san88a}
Sanders, D. B., Soifer, B. T., Elias, J. H., Madore, B. F.,
Matthews, K., Neugebauer, G., \& Scoville, N. Z. 
1988a, \apj, 325, 74

\bibitem[Sanders et~al.(1988b)]{san88b}
Sanders, D. B., Soifer, B. T., Elias, J. H., Neugebauer, G., \& Matthews, K.
1988b, \apj, 328, L35

\bibitem[Sanders et~al.(1989)]{san89}
Sanders, D. B., Phinney, E. S., Neugebauer, G., Soifer, B. T., \& Matthews, K.
1989, \apj, 347, 29

\bibitem[Scoville et~al.(2000)]{sco00}
Scoville, N. Z., et al.
2000, \aj, 119, 991

\bibitem[Smail, Ivison, \& Blain(1997)]{sma97}
Smail, I., Ivison, R. J., \& Blain, A. W.
1997, \apj, 490, L5

\bibitem[Soifer et~al.(1984)]{soi84}
Soifer, B. T., et al.
1984, \apj, 278, L71

\bibitem[Soifer et~al.(1987)]{soi87}
Soifer, B. T., Sanders, D. B., Madore, B. F., Neugebauer, G., Danielson, G. E.,
Elias, J. H., Lonsdale, C. J., \& Rice, W. L.
1987, \apj, 320, 238

\bibitem[Surace et~al.(1998)]{sur98}
Surace, J. A., Sanders, D. B., Vaca, W. D., Veilleux, S., \& Mazzarella, J. M.
1998, \apj, 492, 116

\bibitem[Surace, Sanders, \& Evans(2000)]{sur00}
Surace, J. A., Sanders, D. B., \& Evans, A. S.
2000, \apj, 529, 170

\bibitem[Veilleux et~al.(1995)]{vei95}
Veilleux, S., Kim, D.-C., Sanders, D. B., Mazzarella, J. M., \& Soifer, B. T.
1995, \apjs, 98, 171

\bibitem[Veilleux, Kim, \& Sanders(1999)]{vei99a}
Veilleux, S., Kim, D.-C., \& Sanders, D. B.
1999, \apj, 522, 113

\bibitem[Veilleux, Sanders, \& Kim(1999)]{vei99b}
Veilleux, S., Sanders, D. B., \& Kim, D.-C.
1999, \apj, 522, 139

\bibitem[Zenner \& Lenzen(1993)]{zen93}
Zenner, S., \& Lenzen, R. 1993, \aaps, 101, 363

\bibitem[Zheng et~al.(1999)]{zhe99}
Zhen, Z., Wu, H., Mao, S., Xia, X.-Y., Deng, Z.-G., \& Zou, Z.-L.
1999, \aap, 349, 735

\bibitem[Zhenlong et~al.(1991)]{zhe91}
Zhenlong, Z., Xiaoyong, X., Zugan, D., \& Hongjun, S.
1991, \mnras, 252, 593

\end{thebibliography}
\end{document}